\RequirePackage[2025-19-02]{latexrelease}
\documentclass[aps,amsmath,amssymb,twocolumn,superscriptaddress]{revtex4-2}

\usepackage{graphicx}  % Include figure
\usepackage{textcomp}
\usepackage{gensymb}
\usepackage{color}
\usepackage{placeins}  % Add this in your preamble

\graphicspath{{Plots}}
\usepackage[colorlinks=true, citecolor=blue, linkcolor=blue, urlcolor=blue]{hyperref}
\hypersetup{colorlinks=true, citecolor=blue, linkcolor=blue, urlcolor=blue}
\makeatletter
\AtBeginDocument{%
	\renewcommand{\@biblabel}[1]{[#1]} % Maintain APS citation format
}
\makeatother

\begin{document}
	
	\title{Spin-dependent photovoltage in graphene/MoS$_2$-based field-effect transistors}
	
	\author{K.~Dinar}
	\affiliation{Laboratoire Charles Coulomb (L2C), UMR 5221 CNRS-Université de Montpellier, Montpellier, France.}
	
	\author{J.~Delgado-Notario}
	\affiliation{Nanotechnology Group, USAL-Nanolab, Universidad de Salamanca, Salamanca 37008, Spain.}
	
	\author{C.~Bray}
	\affiliation{Laboratoire Charles Coulomb (L2C), UMR 5221 CNRS-Université de Montpellier, Montpellier, France.}
	
	\author{K.~Maussang}
	\affiliation{IES (UMR 5214), Université de Montpellier, CNRS, Montpellier, France.}
	
	\author{E.~Perez-Martin}
	\affiliation{IES (UMR 5214), Université de Montpellier, CNRS, Montpellier, France.}
	
	\author{B.~Benhamou-Bui}
	\affiliation{Laboratoire Charles Coulomb (L2C), UMR 5221 CNRS-Université de Montpellier, Montpellier, France.}
	
	\author{C.~Consejo}
	\affiliation{Laboratoire Charles Coulomb (L2C), UMR 5221 CNRS-Université de Montpellier, Montpellier, France.}
	
	\author{S.~Ruffenach}
	\affiliation{Laboratoire Charles Coulomb (L2C), UMR 5221 CNRS-Université de Montpellier, Montpellier, France.}
	
	\author{S.S~Krishtopenko}
	\affiliation{Laboratoire Charles Coulomb (L2C), UMR 5221 CNRS-Université de Montpellier, Montpellier, France.}
	
	\author{L.~Bonnet}
	\affiliation{Laboratoire Charles Coulomb (L2C), UMR 5221 CNRS-Université de Montpellier, Montpellier, France.}
	
	\author{M.~Paillet}
	\affiliation{Laboratoire Charles Coulomb (L2C), UMR 5221 CNRS-Université de Montpellier, Montpellier, France.}
	
	\author{J.~Torres}
	\affiliation{Laboratoire Charles Coulomb (L2C), UMR 5221 CNRS-Université de Montpellier, Montpellier, France.}
	
	\author{Y.M.~Meziani}
	\affiliation{Nanotechnology Group, USAL-Nanolab, Universidad de Salamanca, Salamanca 37008, Spain.}
	
	\author{I.~Rozhansky}
	\affiliation{National Graphene Institute, University of Manchester, Manchester M13 9PL, United Kingdom.}
	
	\author{B. Jouault}
	\affiliation{Laboratoire Charles Coulomb (L2C), UMR 5221 CNRS-Université de Montpellier, Montpellier, France.}
	
	\author{S. Nanot}
	\affiliation{Laboratoire Charles Coulomb (L2C), UMR 5221 CNRS-Université de Montpellier, Montpellier, France.}

	\author{F. Teppe$^{*}$}
	\affiliation{Laboratoire Charles Coulomb (L2C), UMR 5221 CNRS-Université de Montpellier, Montpellier, France.}

	\begin{abstract}
		It has recently been shown that Terahertz sensors can effectively detect the spin resonances of Dirac fermions in graphene. The associated photovoltaic measurement technique allows for the investigation of the intrinsic spin-orbit coupling in graphene as well as its topological properties from microwave to Terahertz frequencies. In this work, using graphene/MoS$_2$-based Field-Effect Transistors, we observed a magnetic resonance photovoltage signal in the Gigahertz range that is independent on the gate bias. The dispersion of the associated spin-flip transitions remains intriguingly unaffected by the MoS$_2$ layer. In parallel, the spin-related signal consistently appears as a drop in photovoltage, regardless of the signal's polarity or origin, whether it is due to plasma wave rectification or thermoelectric effects. This behavior is interpreted as a decrease in the system’s spin polarization due to spin-dependent recombination or scattering of photocarriers. Understanding the various photovoltaic signals in highly sensitive Gigahertz/Terahertz sensors paves the way for exploring spin-dependent mechanisms in two-dimensional quantum materials, influenced by proximity effects such as spin-orbit coupling, topology, and magnetism.
	\end{abstract}

	\maketitle  
	
	\begin{center}
		
		\noindent \textbf{*Corresponding author:} \href{mailto:frederic.teppe@umontpellier.fr}{frederic.teppe@umontpellier.fr}
	\end{center}

\section{Introduction}
Over the years, numerous broadband and resonant Terahertz (THz) detectors have been developed, using a wide range of materials and diverse physical mechanisms. These include the Ratchet effect in asymmetric uni- or bi-dimensional structures \cite{monchRatchetEffectSpatially2022, hildTerahertzGigahertzMagnetoratchets2024, abidiTerahertzDetectionAsymmetric2024},thermal effects such as bolometric \cite{jagoMicroscopicOriginBolometric2019, ryzhiiTerahertzBolometricDetectors2023} or thermoelectric responses like the Seebeck effect \cite{bandurinDualOriginRoom2018,vitiThermoelectricGraphenePhotodetectors2021},and Schottky rectification \cite{schlechtEfficientTerahertzRectifier2019, mokharOverviewSemiconductorRectifier2020, shanerEnhancedResponsivityMembrane2007}. Among the different approaches, one particularly notable method was proposed three decades ago by M. Dyakonov and M. Shur \cite{dyakonovShallowWaterAnalogy1993a, dyakonovDetectionMixingFrequency1996a}. They suggested that field-effect transistors (FETs) could detect Terahertz (THz) electromagnetic radiation through rectification of plasmonic oscillations in the two-dimensional (2D) electron gas within the transistor's channel. Since that time, many materials have been explored to design and optimize these broadband or resonant THz detectors \cite{dyakonovGenerationDetectionTerahertz2010},including silicon \cite{taukPlasmaWaveDetection2006, knapPlasmaWaveDetection2004}, GaAs \cite{vekslerDetectionTerahertzRadiation2006a, teppePlasmaWaveResonant2005, teppeRoomtemperaturePlasmaWaves2005},InGaAs \cite{elfatimyResonantVoltagetunableTerahertz2006, boubanga-tombetCurrentDrivenResonant2008}, GaN \cite{elfatimyTerahertzDetectionGaN2006}, HgCdTe \cite{kadykovTerahertzDetectionMagnetic2015b, kadykovTerahertzImagingLandau2016a, ruffenachHgCdTebasedHeterostructuresTerahertz2017}, as well as graphene \cite{vicarelliGrapheneFieldeffectTransistors2012, bandurinResonantTerahertzDetection2018a, caridadRoomTemperaturePlasmonAssistedResonant2024}.

These different THz detectors enable the study of numerous physical properties of the materials used, such as crystalline or structural symmetries \cite{ganichevInterplayRashbaDresselhaus2014}, non-linear carrier dynamics \cite{candussioNonlinearIntensityDependence2021, luUltrafastCarrierDynamics2022}, plasmons and magneto-plasmons \cite{jagerEdgeBulkEffects2001, bandurinCyclotronResonanceOvertones2022}, and Landau level transitions \cite{savchenkoTerahertzCyclotronPhotoconductivity2018}. Interestingly, they can also serve as efficient probes of the high-frequency spin properties of electrons under strong magnetic fields. Based on the ratchet effect in an interdigitated asymmetric top gate device in the presence of a magnetic field, the influence of temperature on different spin and valley splittings in single and bilayer graphene was recently studied \cite{brayTemperaturedependentZerofieldSplittings2022a}. Three electronic spin resonances (ESRs) were indeed identified over a frequency range of 45 GHz to 220 GHz. However, despite a high THz sensitivity, the amplitude of the electron spin resonance (ESR) signal remained relatively weak, i.e., comparable to the signal obtained with the classical electrically detected ESR (EDESR) transport-based technique \cite{maniObservationResistivelyDetected2012a, sichauResonanceMicrowaveMeasurements2019a, singhSublatticeSymmetryBreaking2020a} and the precise origin of this rectification signal was still unclear. To further develop this novel sub-THz photovoltaic ESR technique (PV-ESR), it is therefore necessary to better understand the origin of this ESR signal and to optimize the photovoltaic/photoconductive ESR devices.

In this work, we explore the origin of the ESR signal measured by magneto-photoconductivity in a plasma wave detector based on a monolayer graphene/MoS$_2$ FET. The underlying idea of this structure was to enhance the spin-orbit coupling in the graphene layer through proximity effects with the MoS$_2$ layer as predicted theoretically in \cite{gmitraGrapheneTransitionmetalDichalcogenides2015} through first-principles calculations. This transition metal dichalcogenide substrate has promising potential in this context, theoretically providing substantial enhancements in spin-orbit coupling without affecting graphene's electronic transport properties.

Our experimental results clearly highlight the fact that the PV-ESR signal consistently appears as a drop in photovoltage. This phenomenon is interpreted as a decrease in the system’s spin polarization due to spin-charge conversion, improving our understanding of the origin of the PV-ESR signal. Furthermore, four electron spin resonances are clearly observed and studied as a function of the incident frequency. Interestingly, despite the presence of the MoS$_2$ layer, the observed g-factor and zero field splittings (ZFS) are similar to those observed in pristine monolayer graphene samples that are either CVD-grown or embedded between h-BN. The PV-ESR method thus provides an opportunity to investigate spin-dependent mechanisms in two-dimensional quantum materials influenced by proximity effects, directly within nanoscale FETs. This approach eliminates the need for a cavity or direct contact with the excitation source and operates across a broad frequency range, from a few GHz to several THz.

\section{Sample and characterization}

The plasma wave FET detector studied in our work was fabricated from monolayer graphene and a monolayer of MoS$_2$. The monolayer graphene and MoS$_2$ layers were encapsulated between two layers of hexagonal boron nitride (hBN) using standard dry transfer techniques then deposited on a Si substrate with a 300 nm thick SiO$_2$ layer. The resulting device was patterned in a FET geometry with a channel length of 2 $\mu$m and a width of 14 $\mu$m. A schematic representation and images of the fabricated device are presented in Figures \ref{fig:figure1}(a) and \ref{fig:figure1}(b), with additional details available in the Supplementary Materials \cite{supplement}. When the device is exposed to incident radiation, a signal is generated in the form of a photovoltage $\Delta U_{DS}$ between the drain and source contacts when the source is grounded. An asymmetry is observed between this configuration and the reverse configuration ($\Delta U_{SD}$, with drain grounded), which could potentially be explained by the electron temperature gradient induced by the incident radiation and causing electron diffusion. A detailed analysis of this asymmetry is presented in the Supplementary Materials \cite{supplement}. Since the signal from over-damped plasma wave rectification clearly manifests itself in transconductance \cite{ knapNanometerSizeField2013}, Figure \ref{fig:figure1}(c) represents the transconductances, where a clear charge neutrality point (CNP) is observed around V$_{tg}$ = 0.2 V. The photovoltage measurements under 56 GHz illumination, presented in Figure \ref{fig:figure1}(d), demonstrate a strong photoresponse near the CNP at a temperature of 2 K, which follows, as expected, the shape of the transconductance.
\begin{figure*}[hbtp]
	\centering
	\includegraphics[width=0.9\linewidth]{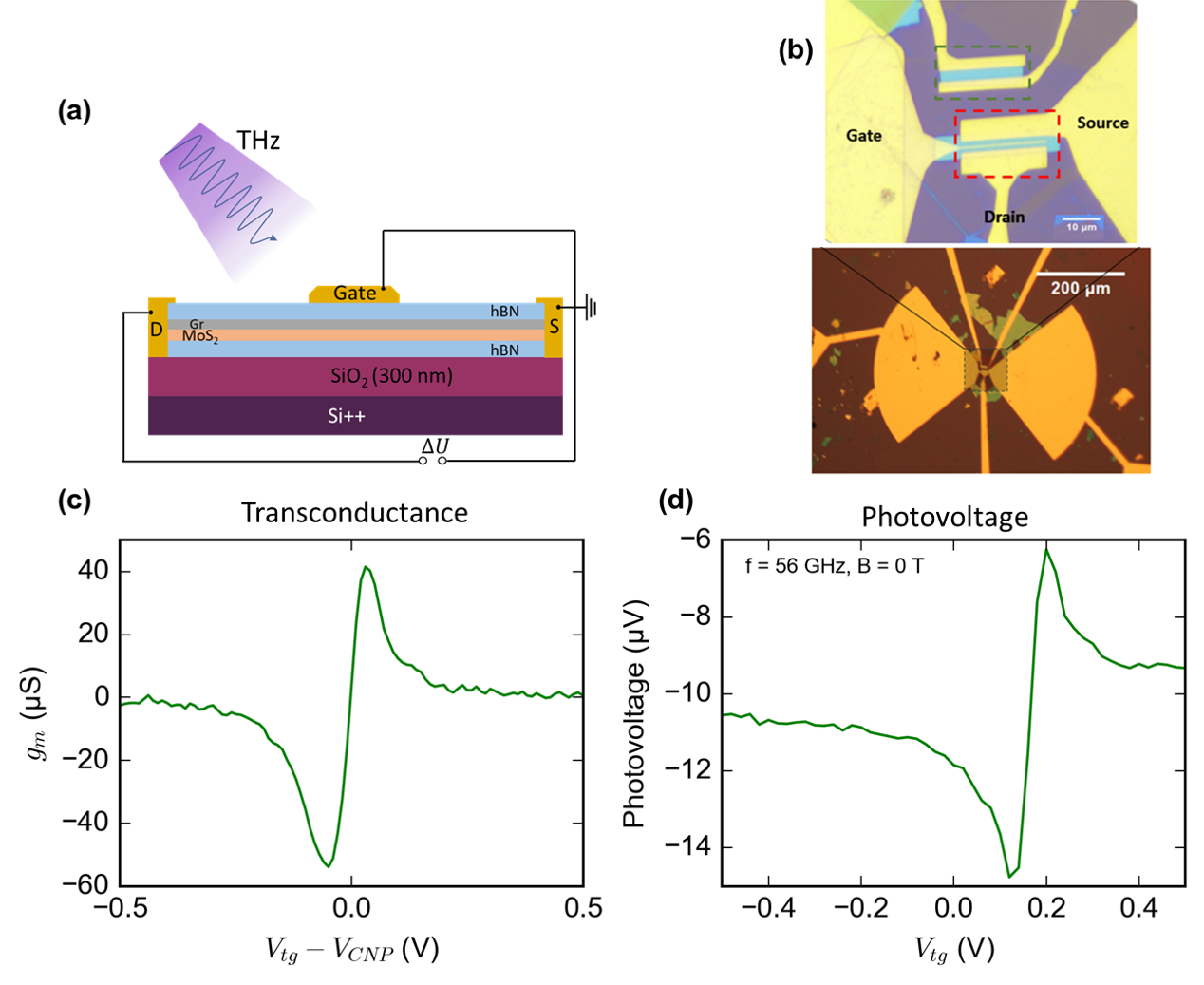}
	\caption{(a) Illustration of the device structure. For characterization, the rectified dc photovoltage ( $\Delta$U) was measured between the source and drain terminals under GHz illumination. (b) Optical image showing the fabricated device with antennas for enhanced GHz coupling. The red dashed square indicates the area of MoS$_2$/Gr. (c) Transconductance (gm) as a function of gate voltage (V$_{tg}$/), measured with a source-drain voltage V$_{SD}$/ = 20 mV and a back-gate voltage V$_{bg}$/ = 0 V, at a temperature T = 2 K. The charge neutrality point (CNP) is observed around V$_{tg}$/ = 0.2 V. (d) Show the photovoltage response under 56 GHz illumination at a temperature of 2 K without a magnetic field. The signal exhibits a high photoresponse around the charge neutrality point (CNP).}
	\label{fig:figure1}
\end{figure*}

\begin{figure*}[hbtp]
	\centering
	\includegraphics[width=0.9\linewidth]{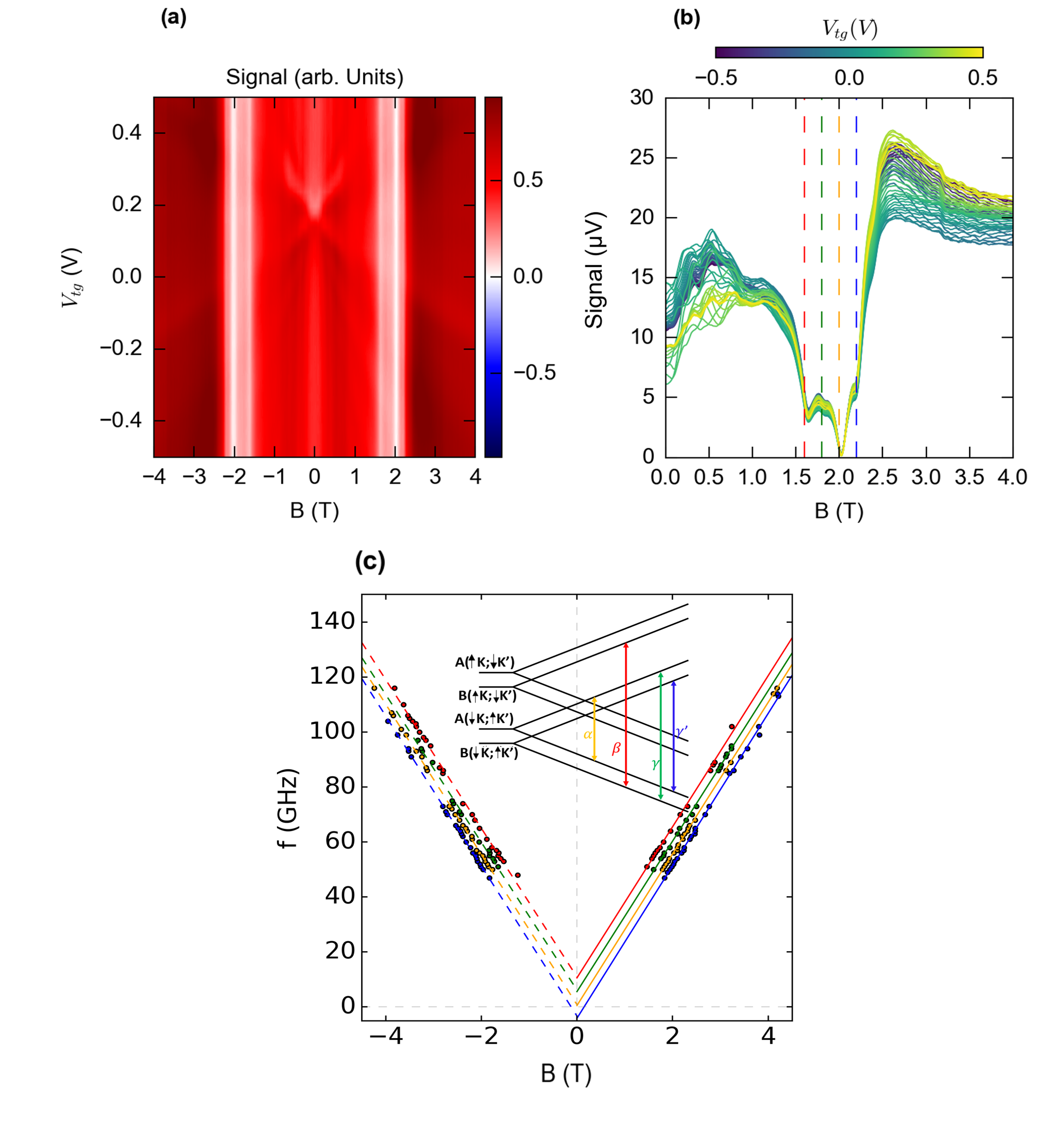}
	\caption{(a) Colormap representation of photovoltage as a function of magnetic field and top gate voltage at 56 GHz. 
		(b) Photovoltage vs. magnetic field for different top gate voltages. Dashed lines of different colors indicate the positions of ESR signals. 
		(c) Position of ESRs as a function of the magnetic field and frequency. Four distinct resonances are marked as $\gamma'$ (blue), $\alpha$ (orange), $\gamma'$ (green), and $\beta$ (red). 
		Solid and dashed lines represent extrapolations for positive and negative magnetic field, respectively. 
		In the inset, a schematic diagram illustrates the energy levels and allowed transitions in graphene under an applied magnetic field. 
		It depicts four energy levels representing different spin and valley states, labeled as A($\uparrow K$; $\downarrow K'$), B($\uparrow K$; $\downarrow K'$), A($\downarrow K$; $\uparrow K'$), and B($\downarrow K$; $\uparrow K'$). 
		The diagram shows the possible resonance transitions indicated by colored arrows: $\alpha$ (orange), $\beta$ (red), $\gamma$ (green), and $\gamma'$ (blue).}
	\label{fig:figure2}
\end{figure*}

When a magnetic field is applied perpendicular to the layers, the source-drain photovoltage signal as a function of the magnetic field reveals Shubnikov-de Haas-like oscillations, which have been observed in several different materials \cite{klimenkoTerahertzResponseInGaAs2010a, boubanga-tombetPlasmaWavesTerahertz2009, boubanga-tombetTerahertzRadiationDetection2009b, kadykovTerahertzDetectionMagnetic2015b, kadykovTerahertzImagingLandau2016a}. The gate voltage-dependent non-resonant signal, especially pronounced in the $\Delta U_{SD}$ configuration (detailed in the supplementary information), arises from the rectification of incident electromagnetic waves via overdamped plasma waves in the transistor channel. The photovoltage signal reflects changes in electronic gas conductivity as the Fermi level oscillates between Landau levels in response to the magnetic field. Figure \ref{fig:figure2}(a) presents a color plot showing the clear evolution of the X feature mean oscillations as a function of magnetic field and gate voltage, alongside field-independent resonance lines. As shown in Figure \ref{fig:figure2}(b), several resonances are visible in the magnetic field range of 1.5 to 2.5 T. These resonances, observed at fixed frequency, are independent of both magnetic field and carrier density, and are attributed to spin-flip transitions between Zeeman-split spin subbands of the conduction band. As seen from Figure and \ref{fig:figure2}(b) and \ref{fig:figure2}(c), the resulting curves contain four electronic spin resonances. These ESRs whose frequency evolves linearly with the magnetic field (see Fig. \ref{fig:figure2}(c)) are studied as a function of the incident excitation frequency up to 115 GHz. The first resonance marked as $\alpha$, extrapolating to zero energy at zero magnetic fields, is attributed to the spin-flip transition between the Zeeman split subbands from the same valley and can be fitted with a Landé g-factor of 1.97 $\pm$ 0.04 calculated from its slope. These ESRs have been first observed using resistively detected ESR measurements in graphene \cite{singhSublatticeSymmetryBreaking2020a}, and later through GHz photovoltage in Ratchet sensors and graphene-based p-n junctions detectors \cite{brayTemperaturedependentZerofieldSplittings2022a}. By extrapolating the other resonances to zero fields, three zero field splittings (ZFSs) are revealed. One of them, named $\beta$, is due to the intrinsic spin-orbit coupling with a frequency of about 11 GHz (corresponding to an energy of 44.5 $\mu$eV, comparable to previously obtained values) and corresponds to the so-called Kane-Mele gap of graphene \cite{brayTemperaturedependentZerofieldSplittings2022a}, while the two others are attributed to sublattice potential with the values also comparable to those reported in the literature \cite{maniObservationResistivelyDetected2012a, sichauResonanceMicrowaveMeasurements2019a, singhSublatticeSymmetryBreaking2020a}, i.e., $\Delta$ $\gamma$’ $\approx$ -4 GHz (corresponding to an energy of about -16.5 5 $\mu$eV), and $\Delta$ $\gamma$ $\approx$ 5.6 GHz (corresponding to 23 5 $\mu$eV).

\section{Discussion}
\subsection{Effect of MoS$_2$}

Surprisingly, roughly the same values are found in several other graphene samples, appearing to be independent of the presence of the MoS$_2$ layer. Even though Raman spectroscopy confirms the presence of a graphene monolayer and indicates good coupling between the graphene and the MoS$_2$ sheets (see Supplementary Materials, Figure SM1 \cite{supplement}), no significant influence of MoS$_2$ is observed. Previous studies using ESR measurements in magneto-transport experiments, have shown that coupling graphene with MoS$_2$ can enhance spin-orbit coupling, leading to a g factor of 1.91 \cite{sharmaElectronSpinResonance2022}. Moreover, similar results were obtained in magic-angle twisted bilayer graphene \cite{morissetteDiracRevivalsDrive2023} and were interpreted as due to some collective excitations linked to the correlated states in magic angle bilayer graphene. However, those results were commented by L. Tiemann et al. \cite{tiemannCommentElectronSpin2022}, arguing that identical resonance features have already been reported in previous works on mono- and few-layer graphene. As an explanation for the experimental results, it was suggested that twisted bilayer graphene may create a slight asymmetry between the two sublattices, similar to that observed in CVD-grown samples in \cite{strenzkeNuclearinducedDephasingSignatures2022}. In our device, interestingly, the sample fabrication method is different, but the ZFS values are similarly consistent with those reported previously (see Fig. \ref{fig:figure2}(b)). Similarly, the g-factor measured in this sample is consistent with that of a graphene sample embedded in h-BN without MoS$_2$ \cite{brayTemperaturedependentZerofieldSplittings2022a}, within the measurement uncertainty (see Fig. SM8 in Supplementary Materials \cite{supplement}). The explanation of this intriguing result requires further research beyond the scope of this work and will be the subject of future studies. Importantly, tight-binding modelling shows that both the magnitude and the type of proximity induced SOC between graphene and MoS$_2$ strongly depends on the Moiré angle between the two materials \cite{naimerTwistangleDependentProximity2021}. This twisted angle is not controlled in our case and is therefore unlikely to correspond to an optimal situation.
\subsection{Origin of the ESR photovoltage}
The PV-ESR signal in graphene-based structures has been previously explained by Bray et al. \cite{brayTemperaturedependentZerofieldSplittings2022a} by the resonant microwave absorption at the spin-flip transition energy, similarly to the commonly used EDESR technique. Alternatively, it was also mentioned that the mechanisms of spin-to-charge conversion, namely spin-dependent scattering, tunnelling, and recombination \cite{mortonEmbracingQuantumLimit2011} may also be at the origin of this PV-ESR signal. Interestingly, as will become clear below, our results highlight that the PV-ESR signal manifests as a significant drop in the photovoltage, almost canceling out at B = 2 T (see Fig. \ref{fig:figure2}(b)). It consistently stays positive, shows a weak dependence on gate voltage, and exhibits minimal distortion from SdH-like oscillations. For the reverse configuration ($\Delta$U$_{SD}$) presented in the supplementary materials \cite{supplement}, the ESR changes sign but always appears as a drop in the photovoltage. To confirm this assumption, photoconductivity measurements were performed by injecting a current into the transistor channel at 56 GHz. Regardless of the baseline signal's polarity or magnitude, which varies with the applied current, the photoconductivity consistently drops toward zero (see Supplementary Materials \cite{supplement}, Figures SM 5,6). Even though it is still impossible to draw conclusions about the exact origin of the signal drop, these results already provide a better understanding of previous findings \cite{kadykovTerahertzImagingLandau2016b, videlierTerahertzPhotovoltaicResponse2011a}, where the ESR signal seemed to change sign erratically. This section therefore presents a non-exhaustive list of probable interpretations of the origin of ESR signal in the photovoltaic or photoconductive regime.

It turns out that this behaviour of the photoconductivity signal was already observed in various systems. Already in the 1970s, Lépine \cite{lepineSpinDependentRecombinationSilicon1972a} measured a decrease in the photoconductivity of silicon samples subjected to powerful microwaves and a magnetic field swept through ESR. This drop of photoconductivity was later observed by several other groups \cite{solomonSpindependentPhotoconductivityNtype1977a, derschRecombinationProcesses$a$Si1983, streetSpindependentPhotoconductivityUndoped1982, schiffSpinPolarizationEffects1981} and explained by a spin-dependent recombination process, in which the recombination rate of the carriers depends on their spin polarization. Consequently, any variation in the spin polarization of the system would affect the recombination time. The steady-state excess carrier concentration $\delta$n is related to the generation rate $G$ and recombination time $\tau$ by $\delta$n = $G \tau$. The change in excess carrier concentration is therefore directly related to the change in photovoltage $\delta$U through the following equation: $\delta$U = $\delta$n $e$ $\mu$, where $e$ is the elementary charge and $\mu$ is the carrier mobility. Consequently, a decrease in $\tau$ leads to a decrease in $\delta$n, resulting in the observed drop in photovoltage.

\begin{figure*}[hbtp]
	\centering
	\includegraphics[width=0.9\linewidth]{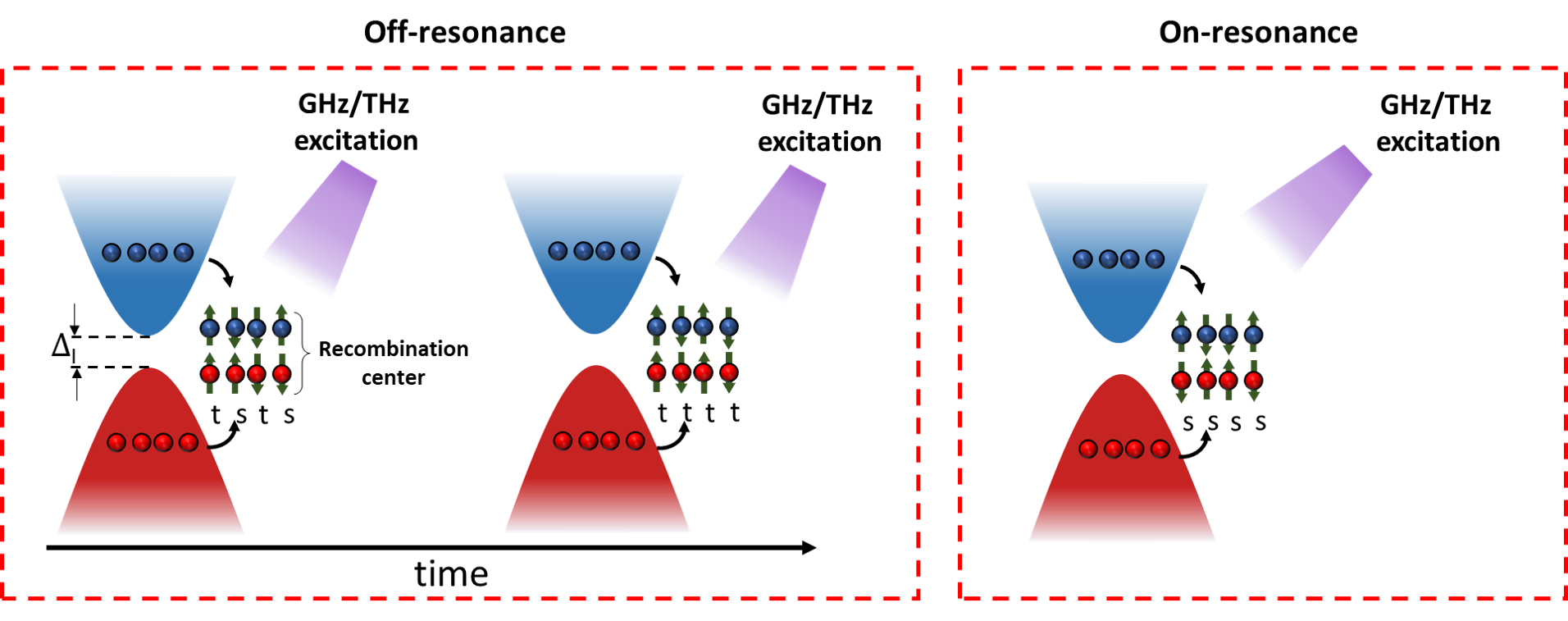}
	\caption{Representation of the Kaplan–Solomon–Mott explanation \cite{solomonSpindependentPhotoconductivityNtype1977a} of spin-dependent recombination adapted to graphene: (a) electron–hole pairs are captured at recombination centers, with singlet pairs rapidly recombining, (b) leading to an accumulation of triplet pairs; (c) electron spin resonance excitation increases the number of singlet pairs, thereby reducing the recombination time.}
	\label{fig:figure3}
\end{figure*}

In more details, Lépine's initial idea was to explain the observed drop of photoconductivity as due to the spin polarization of conduction electrons and paramagnetic recombination centers in the presence of a DC magnetic field. However, this model predicted changes in recombination rates that were several orders of magnitude smaller than those observed experimentally. Alternatively, the Kaplan–Solomon–Mott explanation \cite{kaplanExplanationLargeSpindependent1978a}, based on the Shockley–Read–Hall recombination process and consistent with most experimental observations, suggests that the initial spin polarization arises from the creation of electron-hole pairs trapped at recombination centers before they eventually recombine. An illustration of this idea is sketched in Figure \ref{fig:figure3}. The photoexcited electron-hole pairs in singlet configuration have a shorter lifetime than those induced in triplet configuration. Therefore, the steady-state population of singlet pairs would be smaller, resulting in an excess of triplet configurations that occupy the recombination centers and inhibit further recombination (see Fig. \ref{fig:figure3}(b)). Exciting the system at the ESR frequency returns the recombination centers to a random spin distribution, which contains more singlet states (see Fig. \ref{fig:figure3}(c)), thus reducing the recombination time. Since the concentration of photo-created carriers is a function of the recombination time, and because photoconductivity is proportional to the concentration of photocarriers, a decrease in photoconductivity should occur when the magnetization of the system is reduced. 

Although graphene has an energy gap of approximately 45 $\mu$eV \cite{singhSublatticeSymmetryBreaking2020a, brayTemperaturedependentZerofieldSplittings2022a, sichauResonanceMicrowaveMeasurements2019a}, which is too small for it to exhibit a spin-dependent recombination mechanism identical to the Shockley–Read–Hall process, it may still display similar effects under certain conditions. Our first phenomenological approach is based on the fact that defects and impurities \cite{bhattVariousDefectsGraphene2022}, which create localized states in graphene, can act as spin-dependent recombination centers  \cite{abidReducedGrapheneOxide2018, duttaMagnetizationDueLocalized2015, ghisingGrapheneSpinValves2023}, similar to those in conventional semiconductors. In a phenomenological way the electron-density-dependent mechanism of photovoltage signal suppression can be described as follows. Let us assume two electron species (e.g., spin-up and spin-down electrons or, possibly triplet and singlet configurations of electron-hole pairs) with the sheet densities   n$_1$, n$_2$ determined by a balance between (equal for the two species) generation and recombination rates, so that $G t_1 = n_1; \quad G t_2 = n_2 $. The total density  $n = n_1 + n_2 = G (t_1 + t_2)$ determines the photovoltage signal. In the case of ESR resonance the occupation numbers of the two species are being equalized, which can be described via the following system of rate equations: 
\[
\frac{d n_1}{d t} = G - \frac{n_1}{t_1} - \frac{n_1 - n / 2}{t_{12}}
\]

\[
\frac{d n_2}{d t} = G - \frac{n_2}{t_2} - \frac{n_2 - n / 2}{t_{12}}
\]
where $t_{12}^{-1}$ is a characteristic rate for equilibration of the densities at ESR, this rate is assumed to be fast as compared to the stationary recombination rates: $t_{12} = t_1 , t_2$.

Solving for a stationary state one gets $n = n_0 \frac{4x}{(1 + x)^2}$, where $n_0 = G (t_1 + t_2)$  is the stationary density in the absence of ESR resonance, and $x = \frac{t_2}{t_1}$. At any $x \neq 1$,i.e. if the recombination rates for the two species are unequal, this formula suggests a reduction of the total density due to the equilibration of the occupations of the two species. The physics behind this effect is straightforward and quite general: ESR reduces the lifetime of the longer-living species by redistributing them into the fast recombination channel. However, this model does not fully correspond to our experimental technique. Indeed, unlike the conventional EDESR technique, our photovoltage or photoconductivity technique does not involve resonant absorption, but rather Drude absorption. On the other hand, charge density oscillations are induced by this incident radiation, and the resulting rectification signal arises from the simultaneous modulation of electron velocity and carrier density at the source and gate contacts, respectively. Thus, even if the average electron density remains constant, the density fluctuates due to the oscillatory motion of the electron gas, leading to a nonequilibrium distribution of electrons (with regions of excess and deficit). Excess electrons from regions of high electron density may be trapped at recombination centers and then either recombine or be released due to the nonequilibrium nature of the plasma waves, depending on their lifetime within the traps.

It is generally believed that only spin-dependent recombination should lead to a reduction in photovoltage \cite{cavenettSpinDependentConductivity1979}, even if it has been theoretically demonstrated in silicon transistors \cite{desousaSpindependentScatteringSilicon2009} that spin-dependent scattering mechanisms could also lead to a reduction in conductivity when the process is determined by virtual transitions into doubly occupied donor states. In what follows, we consider an alternative approach that could also result in a reduction of the photovoltage signal during ESR resonance. Indeed, apart from spin-dependant recombination, another mechanism responsible for the observed decrease of the PV-ESR signal might originate from an effect of the ESR on scattering and damping channels without changing the stationary electron density. Electrons can indeed be scattered by neutral donors for example, and since both have spins, the scattering will depend on whether the spins are parallel (triplet) or anti-parallel (singlet), similar to what was previously discussed regarding spin-dependent recombination processes. Let us consider two scattering processes characterized by the time constants $t_1$, $t_2$, characterizing spin-conserving and spin-flip scattering, respectively, the total scattering time $t$ is given by $t^{-1} = t_1^{-1} + t_2^{-1}$. The spin-flip scattering time can be estimated as $t_2^{-1} = \frac{p}{h} \sum_{\mathbf{k}, \mathbf{k}'} \left| V_{\mathbf{k} \mathbf{k}'} \right|^2 \sin^2 q_{\mathbf{k} \mathbf{k}'} D(w)$, where $V_{\mathbf{k} \mathbf{k}'}$ is the matrix element describing the scattering, $\mathbf{k}, \mathbf{k}'$ are the initial and final electron wavevectors $q_{\mathbf{k}, \mathbf{k}'}$ is the angle between them, $D(w) = \frac{1}{p} \frac{\Gamma}{(d_{\text{SO}} - \hbar w)^2 + \Gamma^2}$ is a Lorenzian function describing resonance with a width  $\Gamma$ centered at the spin-orbit splitting $d_{\text{SO}}$ characterizing the energy difference between spin-up and spin-down states. In the context of a simplified phenomenological description, we express the resonant spin-flip time in terms of spin-conserving time through a constant parameter $x$ and the resonant factor $D(w) = t_2^{-1} = t_1^{-1} x D(w)$. The total inverse scattering time would then be $t^{-1} = t_1^{-1} \left( 1 + x D(w) \right)$ exhibiting a resonant maximum at a frequency $w=d_{\text{SO}}$. The resonant increase in the scattering rate would lead to greater damping of the plasma oscillations and reduced conductivity, resulting in a decrease in the photovoltage signal. 

Based on our results, we are unable to distinguish between these two spin-dependent phenomena. Nevertheless, the systematic decrease of the measured signal of spin resonances indicates that the PV-ESR likely stems from a spin-charge conversion phenomenon linked to spin-dependent recombination \cite{lepineSpinDependentRecombinationSilicon1972a} or scattering. When the sample is excited at the ESR frequency, the spin polarization is reduced, which decreases the recombination or scattering time and consequently lowers the photovoltage. Regardless of the origin of the photovoltaic or photoconductive signal — whether it is due to a ratchet effect, a thermoelectric effect, rectification by a p-n junction, or plasma waves — this signal is consistently reduced when ESR is excited. Recent ESR results obtained by Hild and coworkers \cite{hildTerahertzGigahertzMagnetoratchets2024} in ratchet detectors based on graphene metamaterials also showed a reduction in the total signal. Our current results show that the origin of the ESR signal in their work was mistakenly attributed to a Seebeck effect rather than a spin-dependent phenomenon.
\section{Resonances Analysis}
\begin{figure}[hbtp]
	\centering
	\includegraphics[width=0.99\linewidth]{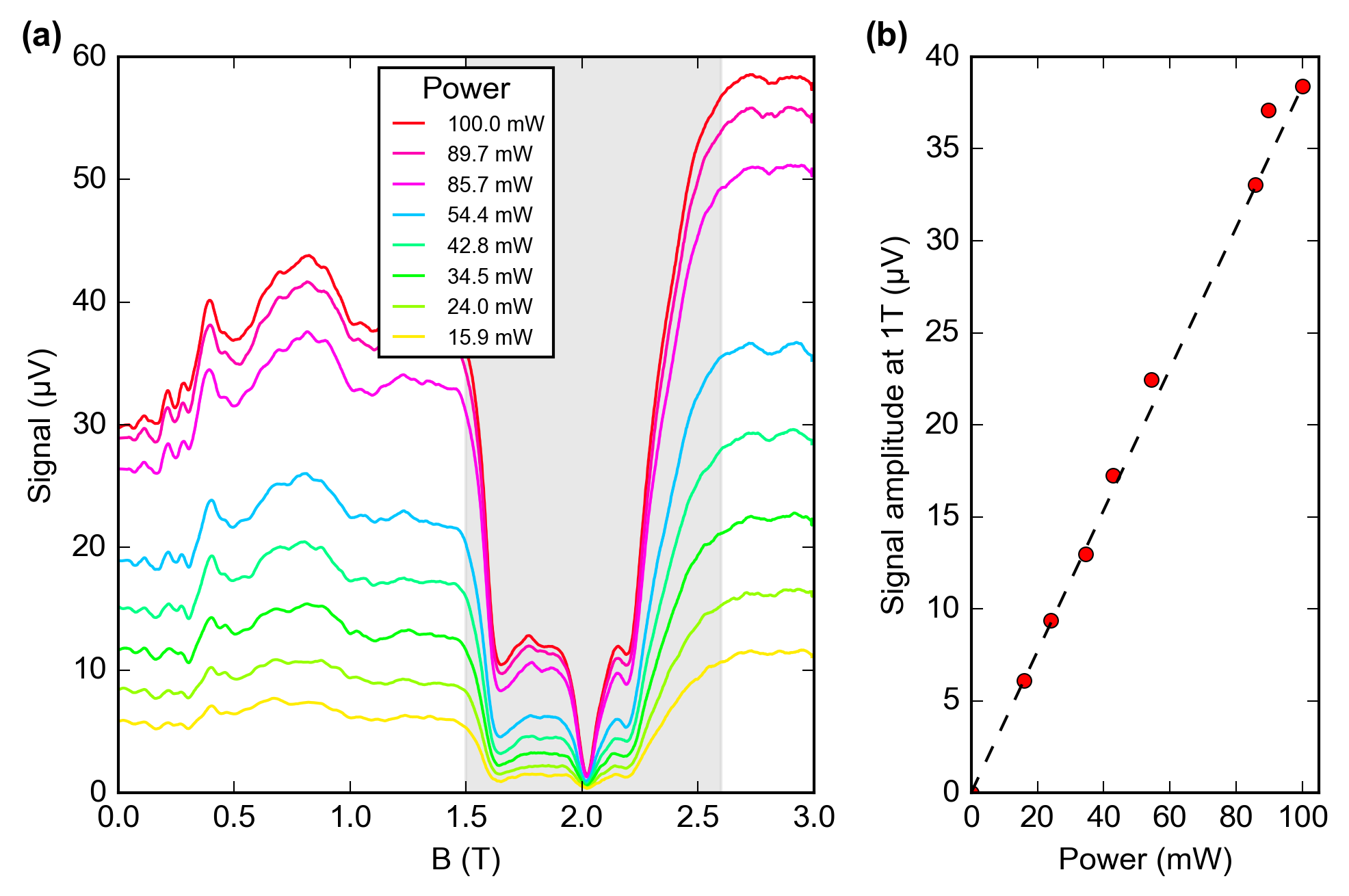}
	\caption{Power dependence of the photovoltage signal at 56 GHz with an injected current of 5 $\mu$A at a temperature of 2 K. (a) Photovoltage signal as a function of magnetic field for different incident powers. The gray shaded area highlights the region between 1.5 and 2.6 T where a significant signal drop is observed. (b) Signal amplitude at 1 T versus incident power, showing a linear relationship. The dashed line serves as an eye guide for the linear trend.}
	\label{fig:figure4}
\end{figure}

\begin{figure}[hbtp]
	\centering
	\includegraphics[width=0.99\linewidth]{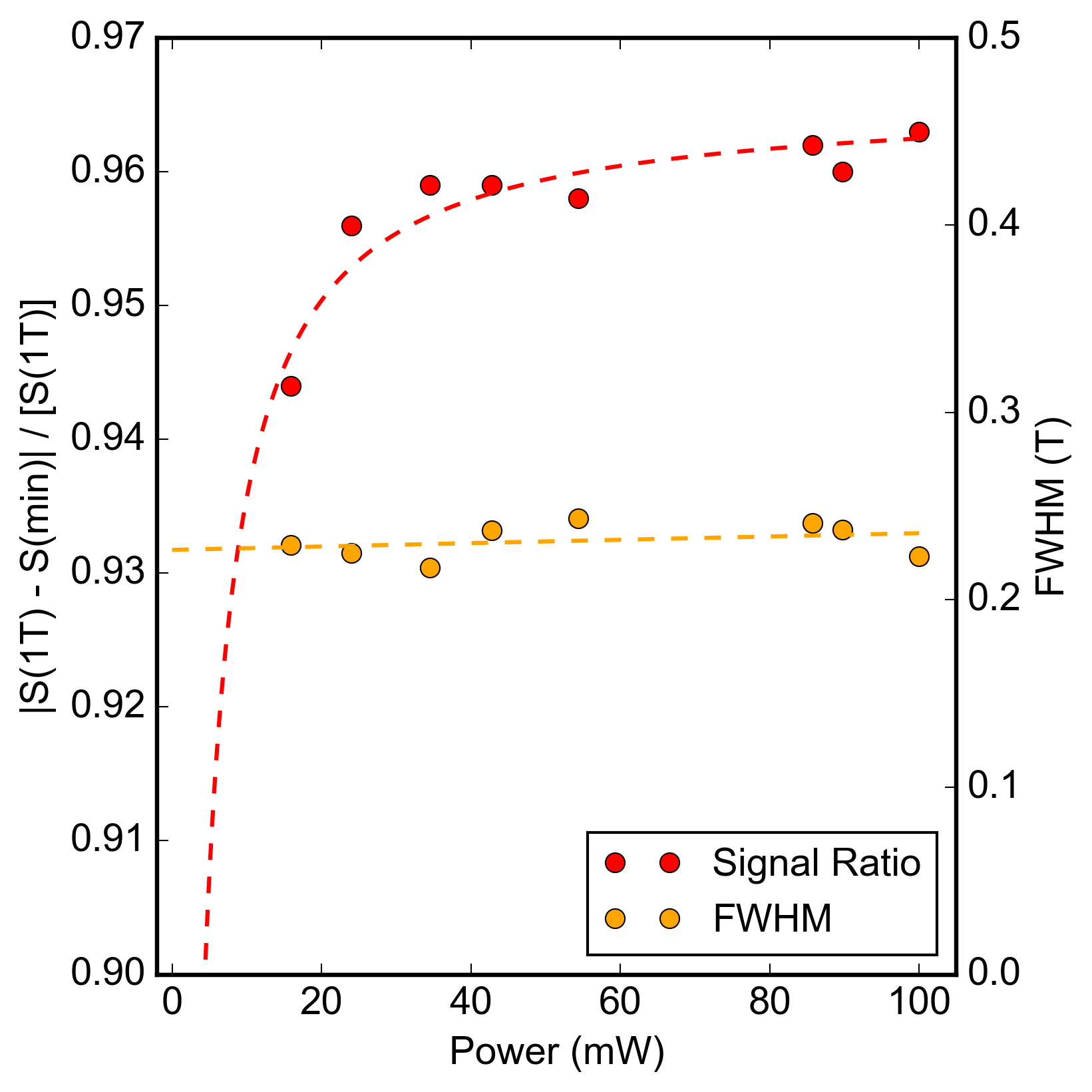}
	\caption{Power dependence of the signal intensity at the center of the resonance. The left y-axis shows the signal intensity \( \frac{|S(1T) - S_{\text{min}}|}{S(1T)} \). The right y-axis displays the full width at half maximum (FWHM) (orange symbols) and its corresponding fit (orange lines). All parameters are plotted against the microwave power in mW. The signal intensity data is fitted using a saturation curve model:
		$y = \frac{aP}{1 + bP}$, where \( P \) is the microwave power and \( a, b \) are fitting parameters. The FWHM is fitted with a power-dependent broadening model:$
		\text{FWHM} = c \sqrt{(1 + aP)}$,	where \( c \) and \( a \) are fitting parameters.}
	\label{fig:figure5}
\end{figure}

In the following, we investigate the effect of incident optical power on the photovoltage signal (see Fig. \ref{fig:figure4}(a)) by considering the two proposed phenomena of spin-dependent scattering and recombination. Firstly, as seen in Figure \ref{fig:figure4}(b), the total signal outside of the ESR, evolves linearly with the incident power. Figure \ref{fig:figure5} then reveals that the ratio of the total signal to the ESR signal under resonant conditions increases at low power and saturates already above 20 mW. 

If we analyse these results considering the effect of spin-dependent scattering as responsible for the photovoltage drop, we can emphasize the role of the electric field generated by the incident radiation at the drain and source contacts, which may influence the spin scattering rate. However, given the weak spin-orbit coupling in graphene, an in-plane electric field should result in only a minor increase in spin depolarization. The Rashba effect may contribute to a slight decrease in the photovoltage, as the accelerating electric field tends to enhance spin depolarization. However, a saturation effect is not expected in this context. This approach, which presents several weaknesses, thus needs a more thorough theoretical assessment.\\	
In the context of spin-dependent recombination, it is possible to analyze the efficiency of spin-flip processes or other resonance-induced phenomena. This is typically measured by the intensity of the ESR signal. With sufficient power, gigahertz excitation generates photo-carriers that reduce the population differences between spin states, potentially saturating the transitions and cancelling the photovoltage. This saturation phenomenon typically occurs due to the spin-lattice relaxation (relaxation time T1), particularly when the concentration of unpaired electrons is low or at low temperatures \cite{l.r.BookReviewPrinciples1961}. The mechanisms by which the spin-lattice interactions can take place all involve interactions of the spin system with phonons \cite{weilElectronParamagneticResonance2007}. The spin-spin relaxation time $\tau_s$ can also be deduced from the full width at half maximum of the ESR signal which is directly related to spin decoherence (See Supplementary Materials \cite{supplement} - Figure SM 7). Using the following equation $ \tau_s = \hbar \left( \frac{2h \Delta f}{\Delta B \, \delta B} \right)^{-1} $ \cite{brayTemperaturedependentZerofieldSplittings2022a}, where \(\hbar\) is the reduced Planck constant, \(\frac{\Delta f}{\Delta B}\) is the slope of the resonance, and \(\delta B\) is the full width at half maximum of the resonance peak, we calculate the spin relaxation time for the \(\alpha\) resonance.

The fits described by the following equations yield key insights into the system's behavior. For the signal intensity, we employ a saturation curve model of the form $aP/(1 + bP)$, where $P$ is the microwave power, and $a$ and $b$ are fitting parameters. The full width at half maximum ($\delta B$) is fitted with a power-dependent broadening model:	$\delta B = c \sqrt{(1 + aP)}$, where $c$ is a fitting parameter. From these fits, we obtain the saturation power (approximately 5–10 mW), maximum signal intensity (around 0.95), and intrinsic linewidth (about 0.23 T, which corresponds to $\tau_s = 12.5$ ps). These spin relaxation values are in good agreement with previous results in the literature \cite{brayTemperaturedependentZerofieldSplittings2022a, singhSublatticeSymmetryBreaking2020a}. The low saturation power indicates high sensitivity of the sensor to GHz excitation. The relatively stable $\delta B$ across the power range implies that inhomogeneous broadening dominates in this system.

\section{Conclusion}
Using a graphene/MoS$_2$ field-effect transistor as a THz/GHz detector, we observed four electronic spin resonances in the photovoltaic signal measured as a function of the magnetic field between 55 and 115 GHz. These resonances are attributed to intrinsic spin-orbit coupling for one, sublattice potential for two others, and to the spin-flip transition between the Zeeman split bands with a Landé g-factor of 1.97 $\pm$ 0.04 for the last one. Interestingly, compared to results from the literature on monolayer graphene samples on various substrates, the ZFS values we measured do not seem to be affected by the presence of the MoS$_2$ layer. This intriguing result will need to be studied in greater detail, both theoretically and experimentally. Additionally, the spin resonances consistently appear as a decrease in the signal, regardless of its origin or polarity. While we cannot definitively determine whether the phenomenon responsible for this effect is spin-dependent scattering or spin-dependent recombination, we interpret the observed signal drop phenomenologically as the system returns to a random spin distribution during ESR excitation. These findings enhance our understanding of photovoltaic detection of the ESR signal in graphene-based GHz/THz detectors. They also highlight the efficiency and potential of these optimized sensors for investigating spin, topological, and proximity phenomena in two-dimensional quantum materials.

\section*{Acknowledgements}
This work was supported by the Terahertz Occitanie Platform. We also acknowledge the Institute for Quantum Technologies in Occitanie (IQO) and the Occitanie Region for their support through a doctoral fellowship. Additional funding was provided by the French Agence Nationale de la Recherche (ANR) under the Jjedi project (JCJC, ANR-18-CE24-0004), the France 2030 program through Equipex+ (Hybat project, ANR-21-ESRE-0026) and PEPR Electronique (Comptera project), as well as by the European Union under the Flag-Era JTC 2019 (DeMeGRaS project, ANR-19-GRF1-0006). J.A.D.-N. thanks the support from Junta de Castilla y León co-funded by FEDER under the Research Grant numbers SA103P23 and from the Universidad de Salamanca for the María Zambrano postdoctoral grant funded by the Next Generation EU Funding for the Requalification of the Spanish University System 2021–23, Spanish Ministry of Universities. Y.M.M acknowledge the support from the Ministry of Science and Innovation (MCIN) and the Spanish State Research Agency (AEI) under grants (PID2021-126483OB-I00).

\section*{Data Availability}

The complete dataset supporting this work is publicly available via the Zenodo repository  \cite{dataset}. \\

\section*{Author contributions}
The experiment was proposed by FT. SN, KD, CB, KM and FT discussed the experimental data and interpretation of the results. The samples were fabricated by KD, JDN and SN. THz photoconductivity experiments were carried out by KD, CB and CC. Characterization measurements were conducted and analysed by BJ, KD, SN, YMM, MP and SR. KM, KD and CB handled the data and prepared the figures. FT wrote the manuscript and all co-authors corrected it. 

\section*{Competing interests}
The authors declare no competing financial interests.

\bibliography{biblio}

\begin{thebibliography}{78}%
\makeatletter
\providecommand \@ifxundefined [1]{%
 \@ifx{#1\undefined}
}%
\providecommand \@ifnum [1]{%
 \ifnum #1\expandafter \@firstoftwo
 \else \expandafter \@secondoftwo
 \fi
}%
\providecommand \@ifx [1]{%
 \ifx #1\expandafter \@firstoftwo
 \else \expandafter \@secondoftwo
 \fi
}%
\providecommand \natexlab [1]{#1}%
\providecommand \enquote  [1]{``#1''}%
\providecommand \bibnamefont  [1]{#1}%
\providecommand \bibfnamefont [1]{#1}%
\providecommand \citenamefont [1]{#1}%
\providecommand \href@noop [0]{\@secondoftwo}%
\providecommand \href [0]{\begingroup \@sanitize@url \@href}%
\providecommand \@href[1]{\@@startlink{#1}\@@href}%
\providecommand \@@href[1]{\endgroup#1\@@endlink}%
\providecommand \@sanitize@url [0]{\catcode `\\12\catcode `\$12\catcode
  `\&12\catcode `\#12\catcode `\^12\catcode `\_12\catcode `\%12\relax}%
\providecommand \@@startlink[1]{}%
\providecommand \@@endlink[0]{}%
\providecommand \url  [0]{\begingroup\@sanitize@url \@url }%
\providecommand \@url [1]{\endgroup\@href {#1}{\urlprefix }}%
\providecommand \urlprefix  [0]{URL }%
\providecommand \Eprint [0]{\href }%
\providecommand \doibase [0]{https://doi.org/}%
\providecommand \selectlanguage [0]{\@gobble}%
\providecommand \bibinfo  [0]{\@secondoftwo}%
\providecommand \bibfield  [0]{\@secondoftwo}%
\providecommand \translation [1]{[#1]}%
\providecommand \BibitemOpen [0]{}%
\providecommand \bibitemStop [0]{}%
\providecommand \bibitemNoStop [0]{.\EOS\space}%
\providecommand \EOS [0]{\spacefactor3000\relax}%
\providecommand \BibitemShut  [1]{\csname bibitem#1\endcsname}%
\let\auto@bib@innerbib\@empty
%</preamble>
\bibitem [{\citenamefont {M{\"o}nch}\ \emph {et~al.}(2022)\citenamefont
  {M{\"o}nch}, \citenamefont {Potashin}, \citenamefont {Lindner}, \citenamefont
  {Yahniuk}, \citenamefont {Golub}, \citenamefont {Kachorovskii}, \citenamefont
  {Bel'kov}, \citenamefont {Huber}, \citenamefont {Watanabe}, \citenamefont
  {Taniguchi}, \citenamefont {Eroms}, \citenamefont {Weiss},\ and\
  \citenamefont {Ganichev}}]{monchRatchetEffectSpatially2022}%
  \BibitemOpen
  \bibfield  {author} {\bibinfo {author} {\bibfnamefont {E.}~\bibnamefont
  {M{\"o}nch}}, \bibinfo {author} {\bibfnamefont {S.~O.}\ \bibnamefont
  {Potashin}}, \bibinfo {author} {\bibfnamefont {K.}~\bibnamefont {Lindner}},
  \bibinfo {author} {\bibfnamefont {I.}~\bibnamefont {Yahniuk}}, \bibinfo
  {author} {\bibfnamefont {L.~E.}\ \bibnamefont {Golub}}, \bibinfo {author}
  {\bibfnamefont {V.~{\relax Yu}.}\ \bibnamefont {Kachorovskii}}, \bibinfo
  {author} {\bibfnamefont {V.~V.}\ \bibnamefont {Bel'kov}}, \bibinfo {author}
  {\bibfnamefont {R.}~\bibnamefont {Huber}}, \bibinfo {author} {\bibfnamefont
  {K.}~\bibnamefont {Watanabe}}, \bibinfo {author} {\bibfnamefont
  {T.}~\bibnamefont {Taniguchi}}, \bibinfo {author} {\bibfnamefont
  {J.}~\bibnamefont {Eroms}}, \bibinfo {author} {\bibfnamefont
  {D.}~\bibnamefont {Weiss}},\ and\ \bibinfo {author} {\bibfnamefont {S.~D.}\
  \bibnamefont {Ganichev}},\ }\href
  {https://doi.org/10.1103/PhysRevB.105.045404} {\bibfield  {journal} {\bibinfo
   {journal} {Physical Review B}\ }\textbf {\bibinfo {volume} {105}},\ \bibinfo
  {pages} {045404} (\bibinfo {year} {2022})}\BibitemShut {NoStop}%
\bibitem [{\citenamefont {Hild}\ \emph {et~al.}(2024)\citenamefont {Hild},
  \citenamefont {M{\"o}nch}, \citenamefont {Golub}, \citenamefont {Dmitriev},
  \citenamefont {Yahniuk}, \citenamefont {Amann}, \citenamefont {Amann},
  \citenamefont {Eroms}, \citenamefont {Wunderlich}, \citenamefont {Weiss},
  \citenamefont {Consejo}, \citenamefont {Bray}, \citenamefont {Maussang},
  \citenamefont {Teppe}, \citenamefont {{Gumenjuk-Sichevska}}, \citenamefont
  {Watanabe}, \citenamefont {Taniguchi},\ and\ \citenamefont
  {Ganichev}}]{hildTerahertzGigahertzMagnetoratchets2024}%
  \BibitemOpen
  \bibfield  {author} {\bibinfo {author} {\bibfnamefont {M.}~\bibnamefont
  {Hild}}, \bibinfo {author} {\bibfnamefont {E.}~\bibnamefont {M{\"o}nch}},
  \bibinfo {author} {\bibfnamefont {L.~E.}\ \bibnamefont {Golub}}, \bibinfo
  {author} {\bibfnamefont {I.~A.}\ \bibnamefont {Dmitriev}}, \bibinfo {author}
  {\bibfnamefont {I.}~\bibnamefont {Yahniuk}}, \bibinfo {author} {\bibfnamefont
  {K.}~\bibnamefont {Amann}}, \bibinfo {author} {\bibfnamefont
  {J.}~\bibnamefont {Amann}}, \bibinfo {author} {\bibfnamefont
  {J.}~\bibnamefont {Eroms}}, \bibinfo {author} {\bibfnamefont
  {J.}~\bibnamefont {Wunderlich}}, \bibinfo {author} {\bibfnamefont
  {D.}~\bibnamefont {Weiss}}, \bibinfo {author} {\bibfnamefont
  {C.}~\bibnamefont {Consejo}}, \bibinfo {author} {\bibfnamefont
  {C.}~\bibnamefont {Bray}}, \bibinfo {author} {\bibfnamefont {K.}~\bibnamefont
  {Maussang}}, \bibinfo {author} {\bibfnamefont {F.}~\bibnamefont {Teppe}},
  \bibinfo {author} {\bibfnamefont {J.}~\bibnamefont {{Gumenjuk-Sichevska}}},
  \bibinfo {author} {\bibfnamefont {K.}~\bibnamefont {Watanabe}}, \bibinfo
  {author} {\bibfnamefont {T.}~\bibnamefont {Taniguchi}},\ and\ \bibinfo
  {author} {\bibfnamefont {S.~D.}\ \bibnamefont {Ganichev}},\ }\href
  {https://doi.org/10.1103/PhysRevB.110.125303} {\bibfield  {journal} {\bibinfo
   {journal} {Physical Review B}\ }\textbf {\bibinfo {volume} {110}},\ \bibinfo
  {pages} {125303} (\bibinfo {year} {2024})}\BibitemShut {NoStop}%
\bibitem [{\citenamefont {Abidi}\ \emph {et~al.}(2024)\citenamefont {Abidi},
  \citenamefont {Khan}, \citenamefont {{Delgado-Notario}}, \citenamefont
  {Cleric{\'o}}, \citenamefont {{Calvo-Gallego}}, \citenamefont {Taniguchi},
  \citenamefont {Watanabe}, \citenamefont {Otsuji}, \citenamefont
  {Vel{\'a}zquez},\ and\ \citenamefont
  {Meziani}}]{abidiTerahertzDetectionAsymmetric2024}%
  \BibitemOpen
  \bibfield  {author} {\bibinfo {author} {\bibfnamefont {E.}~\bibnamefont
  {Abidi}}, \bibinfo {author} {\bibfnamefont {A.}~\bibnamefont {Khan}},
  \bibinfo {author} {\bibfnamefont {J.~A.}\ \bibnamefont {{Delgado-Notario}}},
  \bibinfo {author} {\bibfnamefont {V.}~\bibnamefont {Cleric{\'o}}}, \bibinfo
  {author} {\bibfnamefont {J.}~\bibnamefont {{Calvo-Gallego}}}, \bibinfo
  {author} {\bibfnamefont {T.}~\bibnamefont {Taniguchi}}, \bibinfo {author}
  {\bibfnamefont {K.}~\bibnamefont {Watanabe}}, \bibinfo {author}
  {\bibfnamefont {T.}~\bibnamefont {Otsuji}}, \bibinfo {author} {\bibfnamefont
  {J.~E.}\ \bibnamefont {Vel{\'a}zquez}},\ and\ \bibinfo {author}
  {\bibfnamefont {Y.~M.}\ \bibnamefont {Meziani}},\ }\href
  {https://doi.org/10.3390/nano14040383} {\bibfield  {journal} {\bibinfo
  {journal} {Nanomaterials}\ }\textbf {\bibinfo {volume} {14}},\ \bibinfo
  {pages} {383} (\bibinfo {year} {2024})}\BibitemShut {NoStop}%
\bibitem [{\citenamefont {Jago}\ \emph {et~al.}(2019)\citenamefont {Jago},
  \citenamefont {Malic},\ and\ \citenamefont
  {Wendler}}]{jagoMicroscopicOriginBolometric2019}%
  \BibitemOpen
  \bibfield  {author} {\bibinfo {author} {\bibfnamefont {R.}~\bibnamefont
  {Jago}}, \bibinfo {author} {\bibfnamefont {E.}~\bibnamefont {Malic}},\ and\
  \bibinfo {author} {\bibfnamefont {F.}~\bibnamefont {Wendler}},\ }\href
  {https://doi.org/10.1103/PhysRevB.99.035419} {\bibfield  {journal} {\bibinfo
  {journal} {Physical Review B}\ }\textbf {\bibinfo {volume} {99}},\ \bibinfo
  {pages} {035419} (\bibinfo {year} {2019})}\BibitemShut {NoStop}%
\bibitem [{\citenamefont {Ryzhii}\ \emph {et~al.}(2023)\citenamefont {Ryzhii},
  \citenamefont {Ryzhii}, \citenamefont {Shur}, \citenamefont {Mitin},
  \citenamefont {Tang},\ and\ \citenamefont
  {Otsuji}}]{ryzhiiTerahertzBolometricDetectors2023}%
  \BibitemOpen
  \bibfield  {author} {\bibinfo {author} {\bibfnamefont {M.}~\bibnamefont
  {Ryzhii}}, \bibinfo {author} {\bibfnamefont {V.}~\bibnamefont {Ryzhii}},
  \bibinfo {author} {\bibfnamefont {M.~S.}\ \bibnamefont {Shur}}, \bibinfo
  {author} {\bibfnamefont {V.}~\bibnamefont {Mitin}}, \bibinfo {author}
  {\bibfnamefont {C.}~\bibnamefont {Tang}},\ and\ \bibinfo {author}
  {\bibfnamefont {T.}~\bibnamefont {Otsuji}},\ }\href
  {https://doi.org/10.1063/5.0160899} {\bibfield  {journal} {\bibinfo
  {journal} {Journal of Applied Physics}\ }\textbf {\bibinfo {volume} {134}},\
  \bibinfo {pages} {084501} (\bibinfo {year} {2023})}\BibitemShut {NoStop}%
\bibitem [{\citenamefont {Bandurin}\ \emph
  {et~al.}(2018{\natexlab{a}})\citenamefont {Bandurin}, \citenamefont
  {Gayduchenko}, \citenamefont {Cao}, \citenamefont {Moskotin}, \citenamefont
  {Principi}, \citenamefont {Grigorieva}, \citenamefont {Goltsman},
  \citenamefont {Fedorov},\ and\ \citenamefont
  {Svintsov}}]{bandurinDualOriginRoom2018}%
  \BibitemOpen
  \bibfield  {author} {\bibinfo {author} {\bibfnamefont {D.~A.}\ \bibnamefont
  {Bandurin}}, \bibinfo {author} {\bibfnamefont {I.}~\bibnamefont
  {Gayduchenko}}, \bibinfo {author} {\bibfnamefont {Y.}~\bibnamefont {Cao}},
  \bibinfo {author} {\bibfnamefont {M.}~\bibnamefont {Moskotin}}, \bibinfo
  {author} {\bibfnamefont {A.}~\bibnamefont {Principi}}, \bibinfo {author}
  {\bibfnamefont {I.~V.}\ \bibnamefont {Grigorieva}}, \bibinfo {author}
  {\bibfnamefont {G.}~\bibnamefont {Goltsman}}, \bibinfo {author}
  {\bibfnamefont {G.}~\bibnamefont {Fedorov}},\ and\ \bibinfo {author}
  {\bibfnamefont {D.}~\bibnamefont {Svintsov}},\ }\href
  {https://doi.org/10.1063/1.5018151} {\bibfield  {journal} {\bibinfo
  {journal} {Applied Physics Letters}\ }\textbf {\bibinfo {volume} {112}},\
  \bibinfo {pages} {141101} (\bibinfo {year} {2018}{\natexlab{a}})}\BibitemShut
  {NoStop}%
\bibitem [{\citenamefont {Viti}\ \emph {et~al.}(2021)\citenamefont {Viti},
  \citenamefont {Cadore}, \citenamefont {Yang}, \citenamefont {Vorobiev},
  \citenamefont {Muench}, \citenamefont {Watanabe}, \citenamefont {Taniguchi},
  \citenamefont {Stake}, \citenamefont {Ferrari},\ and\ \citenamefont
  {Vitiello}}]{vitiThermoelectricGraphenePhotodetectors2021}%
  \BibitemOpen
  \bibfield  {author} {\bibinfo {author} {\bibfnamefont {L.}~\bibnamefont
  {Viti}}, \bibinfo {author} {\bibfnamefont {A.~R.}\ \bibnamefont {Cadore}},
  \bibinfo {author} {\bibfnamefont {X.}~\bibnamefont {Yang}}, \bibinfo {author}
  {\bibfnamefont {A.}~\bibnamefont {Vorobiev}}, \bibinfo {author}
  {\bibfnamefont {J.~E.}\ \bibnamefont {Muench}}, \bibinfo {author}
  {\bibfnamefont {K.}~\bibnamefont {Watanabe}}, \bibinfo {author}
  {\bibfnamefont {T.}~\bibnamefont {Taniguchi}}, \bibinfo {author}
  {\bibfnamefont {J.}~\bibnamefont {Stake}}, \bibinfo {author} {\bibfnamefont
  {A.~C.}\ \bibnamefont {Ferrari}},\ and\ \bibinfo {author} {\bibfnamefont
  {M.~S.}\ \bibnamefont {Vitiello}},\ }\href
  {https://doi.org/10.1515/nanoph-2020-0255} {\bibfield  {journal} {\bibinfo
  {journal} {Nanophotonics}\ }\textbf {\bibinfo {volume} {10}},\ \bibinfo
  {pages} {89} (\bibinfo {year} {2021})}\BibitemShut {NoStop}%
\bibitem [{\citenamefont {Schlecht}\ \emph {et~al.}(2019)\citenamefont
  {Schlecht}, \citenamefont {Preu}, \citenamefont {Malzer},\ and\ \citenamefont
  {Weber}}]{schlechtEfficientTerahertzRectifier2019}%
  \BibitemOpen
  \bibfield  {author} {\bibinfo {author} {\bibfnamefont {M.~T.}\ \bibnamefont
  {Schlecht}}, \bibinfo {author} {\bibfnamefont {S.}~\bibnamefont {Preu}},
  \bibinfo {author} {\bibfnamefont {S.}~\bibnamefont {Malzer}},\ and\ \bibinfo
  {author} {\bibfnamefont {H.~B.}\ \bibnamefont {Weber}},\ }\href
  {https://doi.org/10.1038/s41598-019-47606-6} {\bibfield  {journal} {\bibinfo
  {journal} {Scientific Reports}\ }\textbf {\bibinfo {volume} {9}},\ \bibinfo
  {pages} {11205} (\bibinfo {year} {2019})}\BibitemShut {NoStop}%
\bibitem [{\citenamefont {Mokhar}\ \emph {et~al.}(2020)\citenamefont {Mokhar},
  \citenamefont {Kasjoo}, \citenamefont {Juhari},\ and\ \citenamefont
  {Zakaria}}]{mokharOverviewSemiconductorRectifier2020}%
  \BibitemOpen
  \bibfield  {author} {\bibinfo {author} {\bibfnamefont {M.~B.~M.}\
  \bibnamefont {Mokhar}}, \bibinfo {author} {\bibfnamefont {S.~R.}\
  \bibnamefont {Kasjoo}}, \bibinfo {author} {\bibfnamefont {N.~J.}\
  \bibnamefont {Juhari}},\ and\ \bibinfo {author} {\bibfnamefont {N.~F.}\
  \bibnamefont {Zakaria}},\ }\href {https://doi.org/10.1063/1.5142131}
  {\bibfield  {journal} {\bibinfo  {journal} {AIP Conference Proceedings}\
  }\textbf {\bibinfo {volume} {2203}},\ \bibinfo {pages} {020039} (\bibinfo
  {year} {2020})}\BibitemShut {NoStop}%
\bibitem [{\citenamefont {Shaner}\ \emph {et~al.}(2007)\citenamefont {Shaner},
  \citenamefont {Wanke}, \citenamefont {Grine}, \citenamefont {Lyo},
  \citenamefont {Reno},\ and\ \citenamefont
  {Allen}}]{shanerEnhancedResponsivityMembrane2007}%
  \BibitemOpen
  \bibfield  {author} {\bibinfo {author} {\bibfnamefont {E.~A.}\ \bibnamefont
  {Shaner}}, \bibinfo {author} {\bibfnamefont {M.~C.}\ \bibnamefont {Wanke}},
  \bibinfo {author} {\bibfnamefont {A.~D.}\ \bibnamefont {Grine}}, \bibinfo
  {author} {\bibfnamefont {S.~K.}\ \bibnamefont {Lyo}}, \bibinfo {author}
  {\bibfnamefont {J.~L.}\ \bibnamefont {Reno}},\ and\ \bibinfo {author}
  {\bibfnamefont {S.~J.}\ \bibnamefont {Allen}},\ }\href
  {https://doi.org/10.1063/1.2735943} {\bibfield  {journal} {\bibinfo
  {journal} {Applied Physics Letters}\ }\textbf {\bibinfo {volume} {90}},\
  \bibinfo {pages} {181127} (\bibinfo {year} {2007})}\BibitemShut {NoStop}%
\bibitem [{\citenamefont {Dyakonov}\ and\ \citenamefont
  {Shur}(1993)}]{dyakonovShallowWaterAnalogy1993a}%
  \BibitemOpen
  \bibfield  {author} {\bibinfo {author} {\bibfnamefont {M.}~\bibnamefont
  {Dyakonov}}\ and\ \bibinfo {author} {\bibfnamefont {M.}~\bibnamefont
  {Shur}},\ }\href {https://doi.org/10.1103/PhysRevLett.71.2465} {\bibfield
  {journal} {\bibinfo  {journal} {Physical Review Letters}\ }\textbf {\bibinfo
  {volume} {71}},\ \bibinfo {pages} {2465} (\bibinfo {year}
  {1993})}\BibitemShut {NoStop}%
\bibitem [{\citenamefont {Dyakonov}\ and\ \citenamefont
  {Shur}(1996)}]{dyakonovDetectionMixingFrequency1996a}%
  \BibitemOpen
  \bibfield  {author} {\bibinfo {author} {\bibfnamefont {M.}~\bibnamefont
  {Dyakonov}}\ and\ \bibinfo {author} {\bibfnamefont {M.}~\bibnamefont
  {Shur}},\ }\href {https://doi.org/10.1109/16.485650} {\bibfield  {journal}
  {\bibinfo  {journal} {IEEE Transactions on Electron Devices}\ }\textbf
  {\bibinfo {volume} {43}},\ \bibinfo {pages} {380} (\bibinfo {year}
  {1996})}\BibitemShut {NoStop}%
\bibitem [{\citenamefont
  {Dyakonov}(2010)}]{dyakonovGenerationDetectionTerahertz2010}%
  \BibitemOpen
  \bibfield  {author} {\bibinfo {author} {\bibfnamefont {M.~I.}\ \bibnamefont
  {Dyakonov}},\ }\href {https://doi.org/10.1016/j.crhy.2010.05.003} {\bibfield
  {journal} {\bibinfo  {journal} {Comptes Rendus. Physique}\ }\textbf {\bibinfo
  {volume} {11}},\ \bibinfo {pages} {413} (\bibinfo {year} {2010})}\BibitemShut
  {NoStop}%
\bibitem [{\citenamefont {Tauk}\ \emph {et~al.}(2006)\citenamefont {Tauk},
  \citenamefont {Teppe}, \citenamefont {Boubanga}, \citenamefont {Coquillat},
  \citenamefont {Knap}, \citenamefont {Meziani}, \citenamefont {Gallon},
  \citenamefont {Boeuf}, \citenamefont {Skotnicki}, \citenamefont
  {{Fenouillet-Beranger}}, \citenamefont {Maude}, \citenamefont {Rumyantsev},\
  and\ \citenamefont {Shur}}]{taukPlasmaWaveDetection2006}%
  \BibitemOpen
  \bibfield  {author} {\bibinfo {author} {\bibfnamefont {R.}~\bibnamefont
  {Tauk}}, \bibinfo {author} {\bibfnamefont {F.}~\bibnamefont {Teppe}},
  \bibinfo {author} {\bibfnamefont {S.}~\bibnamefont {Boubanga}}, \bibinfo
  {author} {\bibfnamefont {D.}~\bibnamefont {Coquillat}}, \bibinfo {author}
  {\bibfnamefont {W.}~\bibnamefont {Knap}}, \bibinfo {author} {\bibfnamefont
  {Y.~M.}\ \bibnamefont {Meziani}}, \bibinfo {author} {\bibfnamefont
  {C.}~\bibnamefont {Gallon}}, \bibinfo {author} {\bibfnamefont
  {F.}~\bibnamefont {Boeuf}}, \bibinfo {author} {\bibfnamefont
  {T.}~\bibnamefont {Skotnicki}}, \bibinfo {author} {\bibfnamefont
  {C.}~\bibnamefont {{Fenouillet-Beranger}}}, \bibinfo {author} {\bibfnamefont
  {D.~K.}\ \bibnamefont {Maude}}, \bibinfo {author} {\bibfnamefont
  {S.}~\bibnamefont {Rumyantsev}},\ and\ \bibinfo {author} {\bibfnamefont
  {M.~S.}\ \bibnamefont {Shur}},\ }\href {https://doi.org/10.1063/1.2410215}
  {\bibfield  {journal} {\bibinfo  {journal} {Applied Physics Letters}\
  }\textbf {\bibinfo {volume} {89}},\ \bibinfo {pages} {253511} (\bibinfo
  {year} {2006})}\BibitemShut {NoStop}%
\bibitem [{\citenamefont {Knap}\ \emph {et~al.}(2004)\citenamefont {Knap},
  \citenamefont {Teppe}, \citenamefont {Meziani}, \citenamefont {Dyakonova},
  \citenamefont {Lusakowski}, \citenamefont {Boeuf}, \citenamefont {Skotnicki},
  \citenamefont {Maude}, \citenamefont {Rumyantsev},\ and\ \citenamefont
  {Shur}}]{knapPlasmaWaveDetection2004}%
  \BibitemOpen
  \bibfield  {author} {\bibinfo {author} {\bibfnamefont {W.}~\bibnamefont
  {Knap}}, \bibinfo {author} {\bibfnamefont {F.}~\bibnamefont {Teppe}},
  \bibinfo {author} {\bibfnamefont {Y.}~\bibnamefont {Meziani}}, \bibinfo
  {author} {\bibfnamefont {N.}~\bibnamefont {Dyakonova}}, \bibinfo {author}
  {\bibfnamefont {J.}~\bibnamefont {Lusakowski}}, \bibinfo {author}
  {\bibfnamefont {F.}~\bibnamefont {Boeuf}}, \bibinfo {author} {\bibfnamefont
  {T.}~\bibnamefont {Skotnicki}}, \bibinfo {author} {\bibfnamefont
  {D.}~\bibnamefont {Maude}}, \bibinfo {author} {\bibfnamefont
  {S.}~\bibnamefont {Rumyantsev}},\ and\ \bibinfo {author} {\bibfnamefont
  {M.~S.}\ \bibnamefont {Shur}},\ }\href {https://doi.org/10.1063/1.1775034}
  {\bibfield  {journal} {\bibinfo  {journal} {Applied Physics Letters}\
  }\textbf {\bibinfo {volume} {85}},\ \bibinfo {pages} {675} (\bibinfo {year}
  {2004})}\BibitemShut {NoStop}%
\bibitem [{\citenamefont {Veksler}\ \emph {et~al.}(2006)\citenamefont
  {Veksler}, \citenamefont {Teppe}, \citenamefont {Dmitriev}, \citenamefont
  {Kachorovskii}, \citenamefont {Knap},\ and\ \citenamefont
  {Shur}}]{vekslerDetectionTerahertzRadiation2006a}%
  \BibitemOpen
  \bibfield  {author} {\bibinfo {author} {\bibfnamefont {D.}~\bibnamefont
  {Veksler}}, \bibinfo {author} {\bibfnamefont {F.}~\bibnamefont {Teppe}},
  \bibinfo {author} {\bibfnamefont {A.~P.}\ \bibnamefont {Dmitriev}}, \bibinfo
  {author} {\bibfnamefont {V.~{\relax Yu}.}\ \bibnamefont {Kachorovskii}},
  \bibinfo {author} {\bibfnamefont {W.}~\bibnamefont {Knap}},\ and\ \bibinfo
  {author} {\bibfnamefont {M.~S.}\ \bibnamefont {Shur}},\ }\href
  {https://doi.org/10.1103/PhysRevB.73.125328} {\bibfield  {journal} {\bibinfo
  {journal} {Physical Review B}\ }\textbf {\bibinfo {volume} {73}},\ \bibinfo
  {pages} {125328} (\bibinfo {year} {2006})}\BibitemShut {NoStop}%
\bibitem [{\citenamefont {Teppe}\ \emph
  {et~al.}(2005{\natexlab{a}})\citenamefont {Teppe}, \citenamefont {Veksler},
  \citenamefont {Kachorovski}, \citenamefont {Dmitriev}, \citenamefont {Xie},
  \citenamefont {Zhang}, \citenamefont {Rumyantsev}, \citenamefont {Knap},\
  and\ \citenamefont {Shur}}]{teppePlasmaWaveResonant2005}%
  \BibitemOpen
  \bibfield  {author} {\bibinfo {author} {\bibfnamefont {F.}~\bibnamefont
  {Teppe}}, \bibinfo {author} {\bibfnamefont {D.}~\bibnamefont {Veksler}},
  \bibinfo {author} {\bibfnamefont {V.~{\relax Yu}.}\ \bibnamefont
  {Kachorovski}}, \bibinfo {author} {\bibfnamefont {A.~P.}\ \bibnamefont
  {Dmitriev}}, \bibinfo {author} {\bibfnamefont {X.}~\bibnamefont {Xie}},
  \bibinfo {author} {\bibfnamefont {X.-C.}\ \bibnamefont {Zhang}}, \bibinfo
  {author} {\bibfnamefont {S.}~\bibnamefont {Rumyantsev}}, \bibinfo {author}
  {\bibfnamefont {W.}~\bibnamefont {Knap}},\ and\ \bibinfo {author}
  {\bibfnamefont {M.~S.}\ \bibnamefont {Shur}},\ }\href
  {https://doi.org/10.1063/1.1952578} {\bibfield  {journal} {\bibinfo
  {journal} {Applied Physics Letters}\ }\textbf {\bibinfo {volume} {87}},\
  \bibinfo {pages} {022102} (\bibinfo {year} {2005}{\natexlab{a}})}\BibitemShut
  {NoStop}%
\bibitem [{\citenamefont {Teppe}\ \emph
  {et~al.}(2005{\natexlab{b}})\citenamefont {Teppe}, \citenamefont {Knap},
  \citenamefont {Veksler}, \citenamefont {Shur}, \citenamefont {Dmitriev},
  \citenamefont {Kachorovskii},\ and\ \citenamefont
  {Rumyantsev}}]{teppeRoomtemperaturePlasmaWaves2005}%
  \BibitemOpen
  \bibfield  {author} {\bibinfo {author} {\bibfnamefont {F.}~\bibnamefont
  {Teppe}}, \bibinfo {author} {\bibfnamefont {W.}~\bibnamefont {Knap}},
  \bibinfo {author} {\bibfnamefont {D.}~\bibnamefont {Veksler}}, \bibinfo
  {author} {\bibfnamefont {M.~S.}\ \bibnamefont {Shur}}, \bibinfo {author}
  {\bibfnamefont {A.~P.}\ \bibnamefont {Dmitriev}}, \bibinfo {author}
  {\bibfnamefont {V.~{\relax Yu}.}\ \bibnamefont {Kachorovskii}},\ and\
  \bibinfo {author} {\bibfnamefont {S.}~\bibnamefont {Rumyantsev}},\ }\href
  {https://doi.org/10.1063/1.2005394} {\bibfield  {journal} {\bibinfo
  {journal} {Applied Physics Letters}\ }\textbf {\bibinfo {volume} {87}},\
  \bibinfo {pages} {052107} (\bibinfo {year} {2005}{\natexlab{b}})}\BibitemShut
  {NoStop}%
\bibitem [{\citenamefont {El~Fatimy}\ \emph
  {et~al.}(2006{\natexlab{a}})\citenamefont {El~Fatimy}, \citenamefont {Teppe},
  \citenamefont {Dyakonova}, \citenamefont {Knap}, \citenamefont {Seliuta},
  \citenamefont {Valu{\v s}is}, \citenamefont {Shchepetov}, \citenamefont
  {Roelens}, \citenamefont {Bollaert}, \citenamefont {Cappy},\ and\
  \citenamefont {Rumyantsev}}]{elfatimyResonantVoltagetunableTerahertz2006}%
  \BibitemOpen
  \bibfield  {author} {\bibinfo {author} {\bibfnamefont {A.}~\bibnamefont
  {El~Fatimy}}, \bibinfo {author} {\bibfnamefont {F.}~\bibnamefont {Teppe}},
  \bibinfo {author} {\bibfnamefont {N.}~\bibnamefont {Dyakonova}}, \bibinfo
  {author} {\bibfnamefont {W.}~\bibnamefont {Knap}}, \bibinfo {author}
  {\bibfnamefont {D.}~\bibnamefont {Seliuta}}, \bibinfo {author} {\bibfnamefont
  {G.}~\bibnamefont {Valu{\v s}is}}, \bibinfo {author} {\bibfnamefont
  {A.}~\bibnamefont {Shchepetov}}, \bibinfo {author} {\bibfnamefont
  {Y.}~\bibnamefont {Roelens}}, \bibinfo {author} {\bibfnamefont
  {S.}~\bibnamefont {Bollaert}}, \bibinfo {author} {\bibfnamefont
  {A.}~\bibnamefont {Cappy}},\ and\ \bibinfo {author} {\bibfnamefont
  {S.}~\bibnamefont {Rumyantsev}},\ }\href {https://doi.org/10.1063/1.2358816}
  {\bibfield  {journal} {\bibinfo  {journal} {Applied Physics Letters}\
  }\textbf {\bibinfo {volume} {89}},\ \bibinfo {pages} {131926} (\bibinfo
  {year} {2006}{\natexlab{a}})}\BibitemShut {NoStop}%
\bibitem [{\citenamefont {{Boubanga-Tombet}}\ \emph {et~al.}(2008)\citenamefont
  {{Boubanga-Tombet}}, \citenamefont {Teppe}, \citenamefont {Coquillat},
  \citenamefont {Nadar}, \citenamefont {Dyakonova}, \citenamefont {Videlier},
  \citenamefont {Knap}, \citenamefont {Shchepetov}, \citenamefont {Gard{\`e}s},
  \citenamefont {Roelens}, \citenamefont {Bollaert}, \citenamefont {Seliuta},
  \citenamefont {Vadoklis},\ and\ \citenamefont {Valu{\v
  s}is}}]{boubanga-tombetCurrentDrivenResonant2008}%
  \BibitemOpen
  \bibfield  {author} {\bibinfo {author} {\bibfnamefont {S.}~\bibnamefont
  {{Boubanga-Tombet}}}, \bibinfo {author} {\bibfnamefont {F.}~\bibnamefont
  {Teppe}}, \bibinfo {author} {\bibfnamefont {D.}~\bibnamefont {Coquillat}},
  \bibinfo {author} {\bibfnamefont {S.}~\bibnamefont {Nadar}}, \bibinfo
  {author} {\bibfnamefont {N.}~\bibnamefont {Dyakonova}}, \bibinfo {author}
  {\bibfnamefont {H.}~\bibnamefont {Videlier}}, \bibinfo {author}
  {\bibfnamefont {W.}~\bibnamefont {Knap}}, \bibinfo {author} {\bibfnamefont
  {A.}~\bibnamefont {Shchepetov}}, \bibinfo {author} {\bibfnamefont
  {C.}~\bibnamefont {Gard{\`e}s}}, \bibinfo {author} {\bibfnamefont
  {Y.}~\bibnamefont {Roelens}}, \bibinfo {author} {\bibfnamefont
  {S.}~\bibnamefont {Bollaert}}, \bibinfo {author} {\bibfnamefont
  {D.}~\bibnamefont {Seliuta}}, \bibinfo {author} {\bibfnamefont
  {R.}~\bibnamefont {Vadoklis}},\ and\ \bibinfo {author} {\bibfnamefont
  {G.}~\bibnamefont {Valu{\v s}is}},\ }\href
  {https://doi.org/10.1063/1.2936077} {\bibfield  {journal} {\bibinfo
  {journal} {Applied Physics Letters}\ }\textbf {\bibinfo {volume} {92}},\
  \bibinfo {pages} {212101} (\bibinfo {year} {2008})}\BibitemShut {NoStop}%
\bibitem [{\citenamefont {El~Fatimy}\ \emph
  {et~al.}(2006{\natexlab{b}})\citenamefont {El~Fatimy}, \citenamefont
  {Boubanga~Tombet}, \citenamefont {Teppe}, \citenamefont {Knap}, \citenamefont
  {Veksler}, \citenamefont {Rumyantsev}, \citenamefont {Shur}, \citenamefont
  {Pala}, \citenamefont {Gaska}, \citenamefont {Fareed}, \citenamefont {Hu},
  \citenamefont {Seliuta}, \citenamefont {Valusis}, \citenamefont {Gaquiere},
  \citenamefont {Theron},\ and\ \citenamefont
  {Cappy}}]{elfatimyTerahertzDetectionGaN2006}%
  \BibitemOpen
  \bibfield  {author} {\bibinfo {author} {\bibfnamefont {A.}~\bibnamefont
  {El~Fatimy}}, \bibinfo {author} {\bibfnamefont {S.}~\bibnamefont
  {Boubanga~Tombet}}, \bibinfo {author} {\bibfnamefont {F.}~\bibnamefont
  {Teppe}}, \bibinfo {author} {\bibfnamefont {W.}~\bibnamefont {Knap}},
  \bibinfo {author} {\bibfnamefont {D.}~\bibnamefont {Veksler}}, \bibinfo
  {author} {\bibfnamefont {S.}~\bibnamefont {Rumyantsev}}, \bibinfo {author}
  {\bibfnamefont {M.}~\bibnamefont {Shur}}, \bibinfo {author} {\bibfnamefont
  {N.}~\bibnamefont {Pala}}, \bibinfo {author} {\bibfnamefont {R.}~\bibnamefont
  {Gaska}}, \bibinfo {author} {\bibfnamefont {Q.}~\bibnamefont {Fareed}},
  \bibinfo {author} {\bibfnamefont {X.}~\bibnamefont {Hu}}, \bibinfo {author}
  {\bibfnamefont {D.}~\bibnamefont {Seliuta}}, \bibinfo {author} {\bibfnamefont
  {G.}~\bibnamefont {Valusis}}, \bibinfo {author} {\bibfnamefont
  {C.}~\bibnamefont {Gaquiere}}, \bibinfo {author} {\bibfnamefont
  {D.}~\bibnamefont {Theron}},\ and\ \bibinfo {author} {\bibfnamefont
  {A.}~\bibnamefont {Cappy}},\ }\href {https://doi.org/10.1049/el:20062452}
  {\bibfield  {journal} {\bibinfo  {journal} {Electronics Letters}\ }\textbf
  {\bibinfo {volume} {42}},\ \bibinfo {pages} {1342} (\bibinfo {year}
  {2006}{\natexlab{b}})}\BibitemShut {NoStop}%
\bibitem [{\citenamefont {Kadykov}\ \emph {et~al.}(2015)\citenamefont
  {Kadykov}, \citenamefont {Teppe}, \citenamefont {Consejo}, \citenamefont
  {Viti}, \citenamefont {Vitiello}, \citenamefont {Krishtopenko}, \citenamefont
  {Ruffenach}, \citenamefont {Morozov}, \citenamefont {Marcinkiewicz},
  \citenamefont {Desrat}, \citenamefont {Dyakonova}, \citenamefont {Knap},
  \citenamefont {Gavrilenko}, \citenamefont {Mikhailov},\ and\ \citenamefont
  {Dvoretsky}}]{kadykovTerahertzDetectionMagnetic2015b}%
  \BibitemOpen
  \bibfield  {author} {\bibinfo {author} {\bibfnamefont {A.~M.}\ \bibnamefont
  {Kadykov}}, \bibinfo {author} {\bibfnamefont {F.}~\bibnamefont {Teppe}},
  \bibinfo {author} {\bibfnamefont {C.}~\bibnamefont {Consejo}}, \bibinfo
  {author} {\bibfnamefont {L.}~\bibnamefont {Viti}}, \bibinfo {author}
  {\bibfnamefont {M.~S.}\ \bibnamefont {Vitiello}}, \bibinfo {author}
  {\bibfnamefont {S.~S.}\ \bibnamefont {Krishtopenko}}, \bibinfo {author}
  {\bibfnamefont {S.}~\bibnamefont {Ruffenach}}, \bibinfo {author}
  {\bibfnamefont {S.~V.}\ \bibnamefont {Morozov}}, \bibinfo {author}
  {\bibfnamefont {M.}~\bibnamefont {Marcinkiewicz}}, \bibinfo {author}
  {\bibfnamefont {W.}~\bibnamefont {Desrat}}, \bibinfo {author} {\bibfnamefont
  {N.}~\bibnamefont {Dyakonova}}, \bibinfo {author} {\bibfnamefont
  {W.}~\bibnamefont {Knap}}, \bibinfo {author} {\bibfnamefont {V.~I.}\
  \bibnamefont {Gavrilenko}}, \bibinfo {author} {\bibfnamefont {N.~N.}\
  \bibnamefont {Mikhailov}},\ and\ \bibinfo {author} {\bibfnamefont {S.~A.}\
  \bibnamefont {Dvoretsky}},\ }\href {https://doi.org/10.1063/1.4932943}
  {\bibfield  {journal} {\bibinfo  {journal} {Applied Physics Letters}\
  }\textbf {\bibinfo {volume} {107}},\ \bibinfo {pages} {152101} (\bibinfo
  {year} {2015})}\BibitemShut {NoStop}%
\bibitem [{\citenamefont {Kadykov}\ \emph
  {et~al.}(2016{\natexlab{a}})\citenamefont {Kadykov}, \citenamefont {Torres},
  \citenamefont {Krishtopenko}, \citenamefont {Consejo}, \citenamefont
  {Ruffenach}, \citenamefont {Marcinkiewicz}, \citenamefont {But},
  \citenamefont {Knap}, \citenamefont {Morozov}, \citenamefont {Gavrilenko},
  \citenamefont {Mikhailov}, \citenamefont {Dvoretsky},\ and\ \citenamefont
  {Teppe}}]{kadykovTerahertzImagingLandau2016a}%
  \BibitemOpen
  \bibfield  {author} {\bibinfo {author} {\bibfnamefont {A.~M.}\ \bibnamefont
  {Kadykov}}, \bibinfo {author} {\bibfnamefont {J.}~\bibnamefont {Torres}},
  \bibinfo {author} {\bibfnamefont {S.~S.}\ \bibnamefont {Krishtopenko}},
  \bibinfo {author} {\bibfnamefont {C.}~\bibnamefont {Consejo}}, \bibinfo
  {author} {\bibfnamefont {S.}~\bibnamefont {Ruffenach}}, \bibinfo {author}
  {\bibfnamefont {M.}~\bibnamefont {Marcinkiewicz}}, \bibinfo {author}
  {\bibfnamefont {D.}~\bibnamefont {But}}, \bibinfo {author} {\bibfnamefont
  {W.}~\bibnamefont {Knap}}, \bibinfo {author} {\bibfnamefont {S.~V.}\
  \bibnamefont {Morozov}}, \bibinfo {author} {\bibfnamefont {V.~I.}\
  \bibnamefont {Gavrilenko}}, \bibinfo {author} {\bibfnamefont {N.~N.}\
  \bibnamefont {Mikhailov}}, \bibinfo {author} {\bibfnamefont {S.~A.}\
  \bibnamefont {Dvoretsky}},\ and\ \bibinfo {author} {\bibfnamefont
  {F.}~\bibnamefont {Teppe}},\ }\href {https://doi.org/10.1063/1.4955018}
  {\bibfield  {journal} {\bibinfo  {journal} {Applied Physics Letters}\
  }\textbf {\bibinfo {volume} {108}},\ \bibinfo {pages} {262102} (\bibinfo
  {year} {2016}{\natexlab{a}})}\BibitemShut {NoStop}%
\bibitem [{\citenamefont {Ruffenach}\ \emph {et~al.}(2017)\citenamefont
  {Ruffenach}, \citenamefont {Kadykov}, \citenamefont {Rumyantsev},
  \citenamefont {Torres}, \citenamefont {Coquillat}, \citenamefont {But},
  \citenamefont {Krishtopenko}, \citenamefont {Consejo}, \citenamefont {Knap},
  \citenamefont {Winnerl}, \citenamefont {Helm}, \citenamefont {Fadeev},
  \citenamefont {Mikhailov}, \citenamefont {Dvoretskii}, \citenamefont
  {Gavrilenko}, \citenamefont {Morozov},\ and\ \citenamefont
  {Teppe}}]{ruffenachHgCdTebasedHeterostructuresTerahertz2017}%
  \BibitemOpen
  \bibfield  {author} {\bibinfo {author} {\bibfnamefont {S.}~\bibnamefont
  {Ruffenach}}, \bibinfo {author} {\bibfnamefont {A.}~\bibnamefont {Kadykov}},
  \bibinfo {author} {\bibfnamefont {V.~V.}\ \bibnamefont {Rumyantsev}},
  \bibinfo {author} {\bibfnamefont {J.}~\bibnamefont {Torres}}, \bibinfo
  {author} {\bibfnamefont {D.}~\bibnamefont {Coquillat}}, \bibinfo {author}
  {\bibfnamefont {D.}~\bibnamefont {But}}, \bibinfo {author} {\bibfnamefont
  {S.~S.}\ \bibnamefont {Krishtopenko}}, \bibinfo {author} {\bibfnamefont
  {C.}~\bibnamefont {Consejo}}, \bibinfo {author} {\bibfnamefont
  {W.}~\bibnamefont {Knap}}, \bibinfo {author} {\bibfnamefont {S.}~\bibnamefont
  {Winnerl}}, \bibinfo {author} {\bibfnamefont {M.}~\bibnamefont {Helm}},
  \bibinfo {author} {\bibfnamefont {M.~A.}\ \bibnamefont {Fadeev}}, \bibinfo
  {author} {\bibfnamefont {N.~N.}\ \bibnamefont {Mikhailov}}, \bibinfo {author}
  {\bibfnamefont {S.~A.}\ \bibnamefont {Dvoretskii}}, \bibinfo {author}
  {\bibfnamefont {V.~I.}\ \bibnamefont {Gavrilenko}}, \bibinfo {author}
  {\bibfnamefont {S.~V.}\ \bibnamefont {Morozov}},\ and\ \bibinfo {author}
  {\bibfnamefont {F.}~\bibnamefont {Teppe}},\ }\href
  {https://doi.org/10.1063/1.4977781} {\bibfield  {journal} {\bibinfo
  {journal} {APL Materials}\ }\textbf {\bibinfo {volume} {5}},\ \bibinfo
  {pages} {035503} (\bibinfo {year} {2017})}\BibitemShut {NoStop}%
\bibitem [{\citenamefont {Vicarelli}\ \emph {et~al.}(2012)\citenamefont
  {Vicarelli}, \citenamefont {Vitiello}, \citenamefont {Coquillat},
  \citenamefont {Lombardo}, \citenamefont {Ferrari}, \citenamefont {Knap},
  \citenamefont {Polini}, \citenamefont {Pellegrini},\ and\ \citenamefont
  {Tredicucci}}]{vicarelliGrapheneFieldeffectTransistors2012}%
  \BibitemOpen
  \bibfield  {author} {\bibinfo {author} {\bibfnamefont {L.}~\bibnamefont
  {Vicarelli}}, \bibinfo {author} {\bibfnamefont {M.~S.}\ \bibnamefont
  {Vitiello}}, \bibinfo {author} {\bibfnamefont {D.}~\bibnamefont {Coquillat}},
  \bibinfo {author} {\bibfnamefont {A.}~\bibnamefont {Lombardo}}, \bibinfo
  {author} {\bibfnamefont {A.~C.}\ \bibnamefont {Ferrari}}, \bibinfo {author}
  {\bibfnamefont {W.}~\bibnamefont {Knap}}, \bibinfo {author} {\bibfnamefont
  {M.}~\bibnamefont {Polini}}, \bibinfo {author} {\bibfnamefont
  {V.}~\bibnamefont {Pellegrini}},\ and\ \bibinfo {author} {\bibfnamefont
  {A.}~\bibnamefont {Tredicucci}},\ }\href {https://doi.org/10.1038/nmat3417}
  {\bibfield  {journal} {\bibinfo  {journal} {Nature Materials}\ }\textbf
  {\bibinfo {volume} {11}},\ \bibinfo {pages} {865} (\bibinfo {year}
  {2012})}\BibitemShut {NoStop}%
\bibitem [{\citenamefont {Bandurin}\ \emph
  {et~al.}(2018{\natexlab{b}})\citenamefont {Bandurin}, \citenamefont
  {Svintsov}, \citenamefont {Gayduchenko}, \citenamefont {Xu}, \citenamefont
  {Principi}, \citenamefont {Moskotin}, \citenamefont {Tretyakov},
  \citenamefont {Yagodkin}, \citenamefont {Zhukov}, \citenamefont {Taniguchi},
  \citenamefont {Watanabe}, \citenamefont {Grigorieva}, \citenamefont {Polini},
  \citenamefont {Goltsman}, \citenamefont {Geim},\ and\ \citenamefont
  {Fedorov}}]{bandurinResonantTerahertzDetection2018a}%
  \BibitemOpen
  \bibfield  {author} {\bibinfo {author} {\bibfnamefont {D.~A.}\ \bibnamefont
  {Bandurin}}, \bibinfo {author} {\bibfnamefont {D.}~\bibnamefont {Svintsov}},
  \bibinfo {author} {\bibfnamefont {I.}~\bibnamefont {Gayduchenko}}, \bibinfo
  {author} {\bibfnamefont {S.~G.}\ \bibnamefont {Xu}}, \bibinfo {author}
  {\bibfnamefont {A.}~\bibnamefont {Principi}}, \bibinfo {author}
  {\bibfnamefont {M.}~\bibnamefont {Moskotin}}, \bibinfo {author}
  {\bibfnamefont {I.}~\bibnamefont {Tretyakov}}, \bibinfo {author}
  {\bibfnamefont {D.}~\bibnamefont {Yagodkin}}, \bibinfo {author}
  {\bibfnamefont {S.}~\bibnamefont {Zhukov}}, \bibinfo {author} {\bibfnamefont
  {T.}~\bibnamefont {Taniguchi}}, \bibinfo {author} {\bibfnamefont
  {K.}~\bibnamefont {Watanabe}}, \bibinfo {author} {\bibfnamefont {I.~V.}\
  \bibnamefont {Grigorieva}}, \bibinfo {author} {\bibfnamefont
  {M.}~\bibnamefont {Polini}}, \bibinfo {author} {\bibfnamefont {G.~N.}\
  \bibnamefont {Goltsman}}, \bibinfo {author} {\bibfnamefont {A.~K.}\
  \bibnamefont {Geim}},\ and\ \bibinfo {author} {\bibfnamefont
  {G.}~\bibnamefont {Fedorov}},\ }\href
  {https://doi.org/10.1038/s41467-018-07848-w} {\bibfield  {journal} {\bibinfo
  {journal} {Nature Communications}\ }\textbf {\bibinfo {volume} {9}},\
  \bibinfo {pages} {5392} (\bibinfo {year} {2018}{\natexlab{b}})}\BibitemShut
  {NoStop}%
\bibitem [{\citenamefont {Caridad}\ \emph {et~al.}(2024)\citenamefont
  {Caridad}, \citenamefont {Castell{\'o}}, \citenamefont {L{\'o}pez~Baptista},
  \citenamefont {Taniguchi}, \citenamefont {Watanabe}, \citenamefont {Roskos},\
  and\ \citenamefont
  {{Delgado-Notario}}}]{caridadRoomTemperaturePlasmonAssistedResonant2024}%
  \BibitemOpen
  \bibfield  {author} {\bibinfo {author} {\bibfnamefont {J.~M.}\ \bibnamefont
  {Caridad}}, \bibinfo {author} {\bibfnamefont {{\'O}.}~\bibnamefont
  {Castell{\'o}}}, \bibinfo {author} {\bibfnamefont {S.~M.}\ \bibnamefont
  {L{\'o}pez~Baptista}}, \bibinfo {author} {\bibfnamefont {T.}~\bibnamefont
  {Taniguchi}}, \bibinfo {author} {\bibfnamefont {K.}~\bibnamefont {Watanabe}},
  \bibinfo {author} {\bibfnamefont {H.~G.}\ \bibnamefont {Roskos}},\ and\
  \bibinfo {author} {\bibfnamefont {J.~A.}\ \bibnamefont {{Delgado-Notario}}},\
  }\href {https://doi.org/10.1021/acs.nanolett.3c04300} {\bibfield  {journal}
  {\bibinfo  {journal} {Nano Letters}\ }\textbf {\bibinfo {volume} {24}},\
  \bibinfo {pages} {935} (\bibinfo {year} {2024})}\BibitemShut {NoStop}%
\bibitem [{\citenamefont {Ganichev}\ and\ \citenamefont
  {Golub}(2014)}]{ganichevInterplayRashbaDresselhaus2014}%
  \BibitemOpen
  \bibfield  {author} {\bibinfo {author} {\bibfnamefont {S.~D.}\ \bibnamefont
  {Ganichev}}\ and\ \bibinfo {author} {\bibfnamefont {L.~E.}\ \bibnamefont
  {Golub}},\ }\href {https://doi.org/10.1002/pssb.201350261} {\bibfield
  {journal} {\bibinfo  {journal} {physica status solidi (b)}\ }\textbf
  {\bibinfo {volume} {251}},\ \bibinfo {pages} {1801} (\bibinfo {year}
  {2014})}\BibitemShut {NoStop}%
\bibitem [{\citenamefont {Candussio}\ \emph {et~al.}(2021)\citenamefont
  {Candussio}, \citenamefont {Golub}, \citenamefont {Bernreuter}, \citenamefont
  {J{\"o}tten}, \citenamefont {Rockinger}, \citenamefont {Watanabe},
  \citenamefont {Taniguchi}, \citenamefont {Eroms}, \citenamefont {Weiss},\
  and\ \citenamefont {Ganichev}}]{candussioNonlinearIntensityDependence2021}%
  \BibitemOpen
  \bibfield  {author} {\bibinfo {author} {\bibfnamefont {S.}~\bibnamefont
  {Candussio}}, \bibinfo {author} {\bibfnamefont {L.~E.}\ \bibnamefont
  {Golub}}, \bibinfo {author} {\bibfnamefont {S.}~\bibnamefont {Bernreuter}},
  \bibinfo {author} {\bibfnamefont {T.}~\bibnamefont {J{\"o}tten}}, \bibinfo
  {author} {\bibfnamefont {T.}~\bibnamefont {Rockinger}}, \bibinfo {author}
  {\bibfnamefont {K.}~\bibnamefont {Watanabe}}, \bibinfo {author}
  {\bibfnamefont {T.}~\bibnamefont {Taniguchi}}, \bibinfo {author}
  {\bibfnamefont {J.}~\bibnamefont {Eroms}}, \bibinfo {author} {\bibfnamefont
  {D.}~\bibnamefont {Weiss}},\ and\ \bibinfo {author} {\bibfnamefont {S.~D.}\
  \bibnamefont {Ganichev}},\ }\href
  {https://doi.org/10.1103/PhysRevB.104.155404} {\bibfield  {journal} {\bibinfo
   {journal} {Physical Review B}\ }\textbf {\bibinfo {volume} {104}},\ \bibinfo
  {pages} {155404} (\bibinfo {year} {2021})}\BibitemShut {NoStop}%
\bibitem [{\citenamefont {Lu}\ \emph {et~al.}(2022)\citenamefont {Lu},
  \citenamefont {Olvera}, \citenamefont {Turan}, \citenamefont {Seifert},
  \citenamefont {Yardimci}, \citenamefont {Kampfrath}, \citenamefont {Preu},\
  and\ \citenamefont {Jarrahi}}]{luUltrafastCarrierDynamics2022}%
  \BibitemOpen
  \bibfield  {author} {\bibinfo {author} {\bibfnamefont {P.-K.}\ \bibnamefont
  {Lu}}, \bibinfo {author} {\bibfnamefont {A.~d. J.~F.}\ \bibnamefont
  {Olvera}}, \bibinfo {author} {\bibfnamefont {D.}~\bibnamefont {Turan}},
  \bibinfo {author} {\bibfnamefont {T.~S.}\ \bibnamefont {Seifert}}, \bibinfo
  {author} {\bibfnamefont {N.~T.}\ \bibnamefont {Yardimci}}, \bibinfo {author}
  {\bibfnamefont {T.}~\bibnamefont {Kampfrath}}, \bibinfo {author}
  {\bibfnamefont {S.}~\bibnamefont {Preu}},\ and\ \bibinfo {author}
  {\bibfnamefont {M.}~\bibnamefont {Jarrahi}},\ }\href
  {https://doi.org/10.1515/nanoph-2021-0785} {\bibfield  {journal} {\bibinfo
  {journal} {Nanophotonics}\ }\textbf {\bibinfo {volume} {11}},\ \bibinfo
  {pages} {2661} (\bibinfo {year} {2022})}\BibitemShut {NoStop}%
\bibitem [{\citenamefont {Jager}\ \emph {et~al.}(2001)\citenamefont {Jager},
  \citenamefont {Wimmer}, \citenamefont {Lorke}, \citenamefont {Kotthaus},
  \citenamefont {Wegscheider},\ and\ \citenamefont
  {Bichler}}]{jagerEdgeBulkEffects2001}%
  \BibitemOpen
  \bibfield  {author} {\bibinfo {author} {\bibfnamefont {B.~G.~L.}\
  \bibnamefont {Jager}}, \bibinfo {author} {\bibfnamefont {S.}~\bibnamefont
  {Wimmer}}, \bibinfo {author} {\bibfnamefont {A.}~\bibnamefont {Lorke}},
  \bibinfo {author} {\bibfnamefont {J.~P.}\ \bibnamefont {Kotthaus}}, \bibinfo
  {author} {\bibfnamefont {W.}~\bibnamefont {Wegscheider}},\ and\ \bibinfo
  {author} {\bibfnamefont {M.}~\bibnamefont {Bichler}},\ }\href
  {https://doi.org/10.1103/PhysRevB.63.045315} {\bibfield  {journal} {\bibinfo
  {journal} {Physical Review B}\ }\textbf {\bibinfo {volume} {63}},\ \bibinfo
  {pages} {045315} (\bibinfo {year} {2001})}\BibitemShut {NoStop}%
\bibitem [{\citenamefont {Bandurin}\ \emph {et~al.}(2022)\citenamefont
  {Bandurin}, \citenamefont {M{\"o}nch}, \citenamefont {Kapralov},
  \citenamefont {Phinney}, \citenamefont {Lindner}, \citenamefont {Liu},
  \citenamefont {Edgar}, \citenamefont {Dmitriev}, \citenamefont
  {{Jarillo-Herrero}}, \citenamefont {Svintsov},\ and\ \citenamefont
  {Ganichev}}]{bandurinCyclotronResonanceOvertones2022}%
  \BibitemOpen
  \bibfield  {author} {\bibinfo {author} {\bibfnamefont {D.~A.}\ \bibnamefont
  {Bandurin}}, \bibinfo {author} {\bibfnamefont {E.}~\bibnamefont {M{\"o}nch}},
  \bibinfo {author} {\bibfnamefont {K.}~\bibnamefont {Kapralov}}, \bibinfo
  {author} {\bibfnamefont {I.~Y.}\ \bibnamefont {Phinney}}, \bibinfo {author}
  {\bibfnamefont {K.}~\bibnamefont {Lindner}}, \bibinfo {author} {\bibfnamefont
  {S.}~\bibnamefont {Liu}}, \bibinfo {author} {\bibfnamefont {J.~H.}\
  \bibnamefont {Edgar}}, \bibinfo {author} {\bibfnamefont {I.~A.}\ \bibnamefont
  {Dmitriev}}, \bibinfo {author} {\bibfnamefont {P.}~\bibnamefont
  {{Jarillo-Herrero}}}, \bibinfo {author} {\bibfnamefont {D.}~\bibnamefont
  {Svintsov}},\ and\ \bibinfo {author} {\bibfnamefont {S.~D.}\ \bibnamefont
  {Ganichev}},\ }\href {https://doi.org/10.1038/s41567-021-01494-8} {\bibfield
  {journal} {\bibinfo  {journal} {Nature Physics}\ }\textbf {\bibinfo {volume}
  {18}},\ \bibinfo {pages} {462} (\bibinfo {year} {2022})}\BibitemShut
  {NoStop}%
\bibitem [{\citenamefont {Savchenko}\ \emph {et~al.}(2018)\citenamefont
  {Savchenko}, \citenamefont {Kvon}, \citenamefont {Candussio}, \citenamefont
  {Mikhailov}, \citenamefont {Dvoretskii},\ and\ \citenamefont
  {Ganichev}}]{savchenkoTerahertzCyclotronPhotoconductivity2018}%
  \BibitemOpen
  \bibfield  {author} {\bibinfo {author} {\bibfnamefont {M.~L.}\ \bibnamefont
  {Savchenko}}, \bibinfo {author} {\bibfnamefont {Z.~D.}\ \bibnamefont {Kvon}},
  \bibinfo {author} {\bibfnamefont {S.}~\bibnamefont {Candussio}}, \bibinfo
  {author} {\bibfnamefont {N.~N.}\ \bibnamefont {Mikhailov}}, \bibinfo {author}
  {\bibfnamefont {S.~A.}\ \bibnamefont {Dvoretskii}},\ and\ \bibinfo {author}
  {\bibfnamefont {S.~D.}\ \bibnamefont {Ganichev}},\ }\href
  {https://doi.org/10.1134/S0021364018160075} {\bibfield  {journal} {\bibinfo
  {journal} {JETP Letters}\ }\textbf {\bibinfo {volume} {108}},\ \bibinfo
  {pages} {247} (\bibinfo {year} {2018})}\BibitemShut {NoStop}%
\bibitem [{\citenamefont {Bray}\ \emph {et~al.}(2022)\citenamefont {Bray},
  \citenamefont {Maussang}, \citenamefont {Consejo}, \citenamefont
  {{Delgado-Notario}}, \citenamefont {Krishtopenko}, \citenamefont {Yahniuk},
  \citenamefont {Gebert}, \citenamefont {Ruffenach}, \citenamefont {Dinar},
  \citenamefont {Moench}, \citenamefont {Eroms}, \citenamefont {Indykiewicz},
  \citenamefont {Jouault}, \citenamefont {Torres}, \citenamefont {Meziani},
  \citenamefont {Knap}, \citenamefont {Yurgens}, \citenamefont {Ganichev},\
  and\ \citenamefont
  {Teppe}}]{brayTemperaturedependentZerofieldSplittings2022a}%
  \BibitemOpen
  \bibfield  {author} {\bibinfo {author} {\bibfnamefont {C.}~\bibnamefont
  {Bray}}, \bibinfo {author} {\bibfnamefont {K.}~\bibnamefont {Maussang}},
  \bibinfo {author} {\bibfnamefont {C.}~\bibnamefont {Consejo}}, \bibinfo
  {author} {\bibfnamefont {J.~A.}\ \bibnamefont {{Delgado-Notario}}}, \bibinfo
  {author} {\bibfnamefont {S.}~\bibnamefont {Krishtopenko}}, \bibinfo {author}
  {\bibfnamefont {I.}~\bibnamefont {Yahniuk}}, \bibinfo {author} {\bibfnamefont
  {S.}~\bibnamefont {Gebert}}, \bibinfo {author} {\bibfnamefont
  {S.}~\bibnamefont {Ruffenach}}, \bibinfo {author} {\bibfnamefont
  {K.}~\bibnamefont {Dinar}}, \bibinfo {author} {\bibfnamefont
  {E.}~\bibnamefont {Moench}}, \bibinfo {author} {\bibfnamefont
  {J.}~\bibnamefont {Eroms}}, \bibinfo {author} {\bibfnamefont
  {K.}~\bibnamefont {Indykiewicz}}, \bibinfo {author} {\bibfnamefont
  {B.}~\bibnamefont {Jouault}}, \bibinfo {author} {\bibfnamefont
  {J.}~\bibnamefont {Torres}}, \bibinfo {author} {\bibfnamefont {Y.~M.}\
  \bibnamefont {Meziani}}, \bibinfo {author} {\bibfnamefont {W.}~\bibnamefont
  {Knap}}, \bibinfo {author} {\bibfnamefont {A.}~\bibnamefont {Yurgens}},
  \bibinfo {author} {\bibfnamefont {S.~D.}\ \bibnamefont {Ganichev}},\ and\
  \bibinfo {author} {\bibfnamefont {F.}~\bibnamefont {Teppe}},\ }\href
  {https://doi.org/10.1103/PhysRevB.106.245141} {\bibfield  {journal} {\bibinfo
   {journal} {Physical Review B}\ }\textbf {\bibinfo {volume} {106}},\ \bibinfo
  {pages} {245141} (\bibinfo {year} {2022})}\BibitemShut {NoStop}%
\bibitem [{\citenamefont {Mani}\ \emph {et~al.}(2012)\citenamefont {Mani},
  \citenamefont {Hankinson}, \citenamefont {Berger},\ and\ \citenamefont {{de
  Heer}}}]{maniObservationResistivelyDetected2012a}%
  \BibitemOpen
  \bibfield  {author} {\bibinfo {author} {\bibfnamefont {R.~G.}\ \bibnamefont
  {Mani}}, \bibinfo {author} {\bibfnamefont {J.}~\bibnamefont {Hankinson}},
  \bibinfo {author} {\bibfnamefont {C.}~\bibnamefont {Berger}},\ and\ \bibinfo
  {author} {\bibfnamefont {W.~A.}\ \bibnamefont {{de Heer}}},\ }\href
  {https://doi.org/10.1038/ncomms1986} {\bibfield  {journal} {\bibinfo
  {journal} {Nature Communications}\ }\textbf {\bibinfo {volume} {3}},\
  \bibinfo {pages} {996} (\bibinfo {year} {2012})}\BibitemShut {NoStop}%
\bibitem [{\citenamefont {Sichau}\ \emph {et~al.}(2019)\citenamefont {Sichau},
  \citenamefont {Prada}, \citenamefont {Anlauf}, \citenamefont {Lyon},
  \citenamefont {Bosnjak}, \citenamefont {Tiemann},\ and\ \citenamefont
  {Blick}}]{sichauResonanceMicrowaveMeasurements2019a}%
  \BibitemOpen
  \bibfield  {author} {\bibinfo {author} {\bibfnamefont {J.}~\bibnamefont
  {Sichau}}, \bibinfo {author} {\bibfnamefont {M.}~\bibnamefont {Prada}},
  \bibinfo {author} {\bibfnamefont {T.}~\bibnamefont {Anlauf}}, \bibinfo
  {author} {\bibfnamefont {T.~J.}\ \bibnamefont {Lyon}}, \bibinfo {author}
  {\bibfnamefont {B.}~\bibnamefont {Bosnjak}}, \bibinfo {author} {\bibfnamefont
  {L.}~\bibnamefont {Tiemann}},\ and\ \bibinfo {author} {\bibfnamefont {R.~H.}\
  \bibnamefont {Blick}},\ }\href
  {https://doi.org/10.1103/PhysRevLett.122.046403} {\bibfield  {journal}
  {\bibinfo  {journal} {Physical Review Letters}\ }\textbf {\bibinfo {volume}
  {122}},\ \bibinfo {pages} {046403} (\bibinfo {year} {2019})}\BibitemShut
  {NoStop}%
\bibitem [{\citenamefont {Singh}\ \emph {et~al.}(2020)\citenamefont {Singh},
  \citenamefont {Prada}, \citenamefont {Strenzke}, \citenamefont {Bosnjak},
  \citenamefont {Schmirander}, \citenamefont {Tiemann},\ and\ \citenamefont
  {Blick}}]{singhSublatticeSymmetryBreaking2020a}%
  \BibitemOpen
  \bibfield  {author} {\bibinfo {author} {\bibfnamefont {U.~R.}\ \bibnamefont
  {Singh}}, \bibinfo {author} {\bibfnamefont {M.}~\bibnamefont {Prada}},
  \bibinfo {author} {\bibfnamefont {V.}~\bibnamefont {Strenzke}}, \bibinfo
  {author} {\bibfnamefont {B.}~\bibnamefont {Bosnjak}}, \bibinfo {author}
  {\bibfnamefont {T.}~\bibnamefont {Schmirander}}, \bibinfo {author}
  {\bibfnamefont {L.}~\bibnamefont {Tiemann}},\ and\ \bibinfo {author}
  {\bibfnamefont {R.~H.}\ \bibnamefont {Blick}},\ }\href
  {https://doi.org/10.1103/PhysRevB.102.245134} {\bibfield  {journal} {\bibinfo
   {journal} {Physical Review B}\ }\textbf {\bibinfo {volume} {102}},\ \bibinfo
  {pages} {245134} (\bibinfo {year} {2020})}\BibitemShut {NoStop}%
\bibitem [{\citenamefont {Gmitra}\ and\ \citenamefont
  {Fabian}(2015)}]{gmitraGrapheneTransitionmetalDichalcogenides2015}%
  \BibitemOpen
  \bibfield  {author} {\bibinfo {author} {\bibfnamefont {M.}~\bibnamefont
  {Gmitra}}\ and\ \bibinfo {author} {\bibfnamefont {J.}~\bibnamefont
  {Fabian}},\ }\href {https://doi.org/10.1103/PhysRevB.92.155403} {\bibfield
  {journal} {\bibinfo  {journal} {Physical Review B}\ }\textbf {\bibinfo
  {volume} {92}},\ \bibinfo {pages} {155403} (\bibinfo {year}
  {2015})}\BibitemShut {NoStop}%
\bibitem [{sup()}]{supplement}%
  \BibitemOpen
  \href@noop {} {\bibinfo {title} {Supplementary materials}},\ \bibinfo
  {howpublished} {\url{http://link.aps.org/supplemental/PhysRevA}},\ \bibinfo
  {note} {for additional details about sample fabrication, Raman and PL
  characterization, asymmetric photovoltage, magnetic field measurements, DC
  current effects, Lorentzian fitting, and $g$-factor analysis. Also includes
  Refs. \cite{knapNonresonantDetectionTerahertz2002,
  wangOneDimensionalElectricalContact2013a,
  moczkoSymmetrydependentDielectricScreening2023a,
  brayTemperaturedependentZerofieldSplittings2022a,
  vitiEfficientTerahertzDetection2016,
  shabanovOptimalAsymmetryTransistorbased2021,
  caiSensitiveRoomtemperatureTerahertz2014,
  ludwigTerahertzDetectionGraphene2024, lowOriginPhotoresponseBlack2014,
  skoblinThermoelectricEffectsGraphene2017,
  harzheimGeometricallyEnhancedThermoelectric2018,
  annoEnhancementGrapheneThermoelectric2017,
  dollfusThermoelectricEffectsGraphene2015,
  vekslerDetectionTerahertzRadiation2006a,
  sharmaElectronSpinResonance2022}}\BibitemShut {NoStop}%
\bibitem [{\citenamefont {Knap}\ \emph {et~al.}(2013)\citenamefont {Knap},
  \citenamefont {Rumyantsev}, \citenamefont {Vitiello}, \citenamefont
  {Coquillat}, \citenamefont {Blin}, \citenamefont {Dyakonova}, \citenamefont
  {Shur}, \citenamefont {Teppe}, \citenamefont {Tredicucci},\ and\
  \citenamefont {Nagatsuma}}]{knapNanometerSizeField2013}%
  \BibitemOpen
  \bibfield  {author} {\bibinfo {author} {\bibfnamefont {W.}~\bibnamefont
  {Knap}}, \bibinfo {author} {\bibfnamefont {S.}~\bibnamefont {Rumyantsev}},
  \bibinfo {author} {\bibfnamefont {M.~S.}\ \bibnamefont {Vitiello}}, \bibinfo
  {author} {\bibfnamefont {D.}~\bibnamefont {Coquillat}}, \bibinfo {author}
  {\bibfnamefont {S.}~\bibnamefont {Blin}}, \bibinfo {author} {\bibfnamefont
  {N.}~\bibnamefont {Dyakonova}}, \bibinfo {author} {\bibfnamefont
  {M.}~\bibnamefont {Shur}}, \bibinfo {author} {\bibfnamefont {F.}~\bibnamefont
  {Teppe}}, \bibinfo {author} {\bibfnamefont {A.}~\bibnamefont {Tredicucci}},\
  and\ \bibinfo {author} {\bibfnamefont {T.}~\bibnamefont {Nagatsuma}},\ }\href
  {https://doi.org/10.1088/0957-4484/24/21/214002} {\bibfield  {journal}
  {\bibinfo  {journal} {Nanotechnology}\ }\textbf {\bibinfo {volume} {24}},\
  \bibinfo {pages} {214002} (\bibinfo {year} {2013})}\BibitemShut {NoStop}%
\bibitem [{\citenamefont {Klimenko}\ \emph {et~al.}(2010)\citenamefont
  {Klimenko}, \citenamefont {Mityagin}, \citenamefont {Videlier}, \citenamefont
  {Teppe}, \citenamefont {Dyakonova}, \citenamefont {Consejo}, \citenamefont
  {Bollaert}, \citenamefont {Murzin},\ and\ \citenamefont
  {Knap}}]{klimenkoTerahertzResponseInGaAs2010a}%
  \BibitemOpen
  \bibfield  {author} {\bibinfo {author} {\bibfnamefont {O.~A.}\ \bibnamefont
  {Klimenko}}, \bibinfo {author} {\bibfnamefont {{\relax Yu}.~A.}\ \bibnamefont
  {Mityagin}}, \bibinfo {author} {\bibfnamefont {H.}~\bibnamefont {Videlier}},
  \bibinfo {author} {\bibfnamefont {F.}~\bibnamefont {Teppe}}, \bibinfo
  {author} {\bibfnamefont {N.~V.}\ \bibnamefont {Dyakonova}}, \bibinfo {author}
  {\bibfnamefont {C.}~\bibnamefont {Consejo}}, \bibinfo {author} {\bibfnamefont
  {S.}~\bibnamefont {Bollaert}}, \bibinfo {author} {\bibfnamefont {V.~N.}\
  \bibnamefont {Murzin}},\ and\ \bibinfo {author} {\bibfnamefont
  {W.}~\bibnamefont {Knap}},\ }\href {https://doi.org/10.1063/1.3462072}
  {\bibfield  {journal} {\bibinfo  {journal} {Applied Physics Letters}\
  }\textbf {\bibinfo {volume} {97}},\ \bibinfo {pages} {022111} (\bibinfo
  {year} {2010})}\BibitemShut {NoStop}%
\bibitem [{\citenamefont {{Boubanga-Tombet}}\ \emph
  {et~al.}(2009{\natexlab{a}})\citenamefont {{Boubanga-Tombet}}, \citenamefont
  {Teppe}, \citenamefont {Knap}, \citenamefont {Karpierz}, \citenamefont
  {Lusakowski}, \citenamefont {Grynberg}, \citenamefont {Dyakonov},
  \citenamefont {Lyonnet},\ and\ \citenamefont
  {Peiris}}]{boubanga-tombetPlasmaWavesTerahertz2009}%
  \BibitemOpen
  \bibfield  {author} {\bibinfo {author} {\bibfnamefont {S.}~\bibnamefont
  {{Boubanga-Tombet}}}, \bibinfo {author} {\bibfnamefont {F.}~\bibnamefont
  {Teppe}}, \bibinfo {author} {\bibfnamefont {W.}~\bibnamefont {Knap}},
  \bibinfo {author} {\bibfnamefont {K.}~\bibnamefont {Karpierz}}, \bibinfo
  {author} {\bibfnamefont {J.}~\bibnamefont {Lusakowski}}, \bibinfo {author}
  {\bibfnamefont {M.}~\bibnamefont {Grynberg}}, \bibinfo {author}
  {\bibfnamefont {M.~I.}\ \bibnamefont {Dyakonov}}, \bibinfo {author}
  {\bibfnamefont {J.}~\bibnamefont {Lyonnet}},\ and\ \bibinfo {author}
  {\bibfnamefont {J.~M.}\ \bibnamefont {Peiris}},\ }\href
  {https://doi.org/10.1088/1742-6596/193/1/012083} {\bibfield  {journal}
  {\bibinfo  {journal} {Journal of Physics: Conference Series}\ }\textbf
  {\bibinfo {volume} {193}},\ \bibinfo {pages} {012083} (\bibinfo {year}
  {2009}{\natexlab{a}})}\BibitemShut {NoStop}%
\bibitem [{\citenamefont {{Boubanga-Tombet}}\ \emph
  {et~al.}(2009{\natexlab{b}})\citenamefont {{Boubanga-Tombet}}, \citenamefont
  {Sakowicz}, \citenamefont {Coquillat}, \citenamefont {Teppe}, \citenamefont
  {Knap}, \citenamefont {Dyakonov}, \citenamefont {Karpierz}, \citenamefont
  {{\L}usakowski},\ and\ \citenamefont
  {Grynberg}}]{boubanga-tombetTerahertzRadiationDetection2009b}%
  \BibitemOpen
  \bibfield  {author} {\bibinfo {author} {\bibfnamefont {S.}~\bibnamefont
  {{Boubanga-Tombet}}}, \bibinfo {author} {\bibfnamefont {M.}~\bibnamefont
  {Sakowicz}}, \bibinfo {author} {\bibfnamefont {D.}~\bibnamefont {Coquillat}},
  \bibinfo {author} {\bibfnamefont {F.}~\bibnamefont {Teppe}}, \bibinfo
  {author} {\bibfnamefont {W.}~\bibnamefont {Knap}}, \bibinfo {author}
  {\bibfnamefont {M.~I.}\ \bibnamefont {Dyakonov}}, \bibinfo {author}
  {\bibfnamefont {K.}~\bibnamefont {Karpierz}}, \bibinfo {author}
  {\bibfnamefont {J.}~\bibnamefont {{\L}usakowski}},\ and\ \bibinfo {author}
  {\bibfnamefont {M.}~\bibnamefont {Grynberg}},\ }\href
  {https://doi.org/10.1063/1.3207886} {\bibfield  {journal} {\bibinfo
  {journal} {Applied Physics Letters}\ }\textbf {\bibinfo {volume} {95}},\
  \bibinfo {pages} {072106} (\bibinfo {year} {2009}{\natexlab{b}})}\BibitemShut
  {NoStop}%
\bibitem [{\citenamefont {Sharma}\ \emph {et~al.}(2022)\citenamefont {Sharma},
  \citenamefont {Zhao}, \citenamefont {Tiemann}, \citenamefont {Prada},
  \citenamefont {Pandey}, \citenamefont {Stierle},\ and\ \citenamefont
  {Blick}}]{sharmaElectronSpinResonance2022}%
  \BibitemOpen
  \bibfield  {author} {\bibinfo {author} {\bibfnamefont {C.~H.}\ \bibnamefont
  {Sharma}}, \bibinfo {author} {\bibfnamefont {P.}~\bibnamefont {Zhao}},
  \bibinfo {author} {\bibfnamefont {L.}~\bibnamefont {Tiemann}}, \bibinfo
  {author} {\bibfnamefont {M.}~\bibnamefont {Prada}}, \bibinfo {author}
  {\bibfnamefont {A.~D.}\ \bibnamefont {Pandey}}, \bibinfo {author}
  {\bibfnamefont {A.}~\bibnamefont {Stierle}},\ and\ \bibinfo {author}
  {\bibfnamefont {R.~H.}\ \bibnamefont {Blick}},\ }\href
  {https://doi.org/10.1063/5.0077077} {\bibfield  {journal} {\bibinfo
  {journal} {AIP Advances}\ }\textbf {\bibinfo {volume} {12}},\ \bibinfo
  {pages} {035111} (\bibinfo {year} {2022})}\BibitemShut {NoStop}%
\bibitem [{\citenamefont {Morissette}\ \emph {et~al.}(2023)\citenamefont
  {Morissette}, \citenamefont {Lin}, \citenamefont {Sun}, \citenamefont
  {Zhang}, \citenamefont {Liu}, \citenamefont {Rhodes}, \citenamefont
  {Watanabe}, \citenamefont {Taniguchi}, \citenamefont {Hone}, \citenamefont
  {Pollanen}, \citenamefont {Scheurer}, \citenamefont {Lilly}, \citenamefont
  {Mounce},\ and\ \citenamefont {Li}}]{morissetteDiracRevivalsDrive2023}%
  \BibitemOpen
  \bibfield  {author} {\bibinfo {author} {\bibfnamefont {E.}~\bibnamefont
  {Morissette}}, \bibinfo {author} {\bibfnamefont {J.-X.}\ \bibnamefont {Lin}},
  \bibinfo {author} {\bibfnamefont {D.}~\bibnamefont {Sun}}, \bibinfo {author}
  {\bibfnamefont {L.}~\bibnamefont {Zhang}}, \bibinfo {author} {\bibfnamefont
  {S.}~\bibnamefont {Liu}}, \bibinfo {author} {\bibfnamefont {D.}~\bibnamefont
  {Rhodes}}, \bibinfo {author} {\bibfnamefont {K.}~\bibnamefont {Watanabe}},
  \bibinfo {author} {\bibfnamefont {T.}~\bibnamefont {Taniguchi}}, \bibinfo
  {author} {\bibfnamefont {J.}~\bibnamefont {Hone}}, \bibinfo {author}
  {\bibfnamefont {J.}~\bibnamefont {Pollanen}}, \bibinfo {author}
  {\bibfnamefont {M.~S.}\ \bibnamefont {Scheurer}}, \bibinfo {author}
  {\bibfnamefont {M.}~\bibnamefont {Lilly}}, \bibinfo {author} {\bibfnamefont
  {A.}~\bibnamefont {Mounce}},\ and\ \bibinfo {author} {\bibfnamefont
  {J.~I.~A.}\ \bibnamefont {Li}},\ }\href
  {https://doi.org/10.1038/s41567-023-02060-0} {\bibfield  {journal} {\bibinfo
  {journal} {Nature Physics}\ }\textbf {\bibinfo {volume} {19}},\ \bibinfo
  {pages} {1156} (\bibinfo {year} {2023})}\BibitemShut {NoStop}%
\bibitem [{\citenamefont {Tiemann}\ \emph {et~al.}(2022)\citenamefont
  {Tiemann}, \citenamefont {Prada}, \citenamefont {Strenzke},\ and\
  \citenamefont {Blick}}]{tiemannCommentElectronSpin2022}%
  \BibitemOpen
  \bibfield  {author} {\bibinfo {author} {\bibfnamefont {L.}~\bibnamefont
  {Tiemann}}, \bibinfo {author} {\bibfnamefont {M.}~\bibnamefont {Prada}},
  \bibinfo {author} {\bibfnamefont {V.}~\bibnamefont {Strenzke}},\ and\
  \bibinfo {author} {\bibfnamefont {R.~H.}\ \bibnamefont {Blick}},\ }\href
  {https://doi.org/10.48550/arXiv.2206.14628} {\bibinfo {title} {Comment on
  "{{Electron}} spin resonance and collective excitations in magic-angle
  twisted bilayer graphene"}} (\bibinfo {year} {2022}),\ \Eprint
  {https://arxiv.org/abs/2206.14628} {arXiv:2206.14628 [cond-mat]} \BibitemShut
  {NoStop}%
\bibitem [{\citenamefont {Strenzke}\ \emph {et~al.}(2022)\citenamefont
  {Strenzke}, \citenamefont {Meyer}, \citenamefont {{Grandt-Ionita}},
  \citenamefont {Prada}, \citenamefont {Kim}, \citenamefont {Heilmann},
  \citenamefont {Lopes}, \citenamefont {Tiemann},\ and\ \citenamefont
  {Blick}}]{strenzkeNuclearinducedDephasingSignatures2022}%
  \BibitemOpen
  \bibfield  {author} {\bibinfo {author} {\bibfnamefont {V.}~\bibnamefont
  {Strenzke}}, \bibinfo {author} {\bibfnamefont {J.~M.}\ \bibnamefont {Meyer}},
  \bibinfo {author} {\bibfnamefont {I.}~\bibnamefont {{Grandt-Ionita}}},
  \bibinfo {author} {\bibfnamefont {M.}~\bibnamefont {Prada}}, \bibinfo
  {author} {\bibfnamefont {H.-S.}\ \bibnamefont {Kim}}, \bibinfo {author}
  {\bibfnamefont {M.}~\bibnamefont {Heilmann}}, \bibinfo {author}
  {\bibfnamefont {J.~M.~J.}\ \bibnamefont {Lopes}}, \bibinfo {author}
  {\bibfnamefont {L.}~\bibnamefont {Tiemann}},\ and\ \bibinfo {author}
  {\bibfnamefont {R.~H.}\ \bibnamefont {Blick}},\ }\href
  {https://doi.org/10.1103/PhysRevB.105.144303} {\bibfield  {journal} {\bibinfo
   {journal} {Physical Review B}\ }\textbf {\bibinfo {volume} {105}},\ \bibinfo
  {pages} {144303} (\bibinfo {year} {2022})}\BibitemShut {NoStop}%
\bibitem [{\citenamefont {Naimer}\ \emph {et~al.}(2021)\citenamefont {Naimer},
  \citenamefont {Zollner}, \citenamefont {Gmitra},\ and\ \citenamefont
  {Fabian}}]{naimerTwistangleDependentProximity2021}%
  \BibitemOpen
  \bibfield  {author} {\bibinfo {author} {\bibfnamefont {T.}~\bibnamefont
  {Naimer}}, \bibinfo {author} {\bibfnamefont {K.}~\bibnamefont {Zollner}},
  \bibinfo {author} {\bibfnamefont {M.}~\bibnamefont {Gmitra}},\ and\ \bibinfo
  {author} {\bibfnamefont {J.}~\bibnamefont {Fabian}},\ }\href
  {https://doi.org/10.1103/PhysRevB.104.195156} {\bibfield  {journal} {\bibinfo
   {journal} {Physical Review B}\ }\textbf {\bibinfo {volume} {104}},\ \bibinfo
  {pages} {195156} (\bibinfo {year} {2021})}\BibitemShut {NoStop}%
\bibitem [{\citenamefont {Morton}\ \emph {et~al.}(2011)\citenamefont {Morton},
  \citenamefont {McCamey}, \citenamefont {Eriksson},\ and\ \citenamefont
  {Lyon}}]{mortonEmbracingQuantumLimit2011}%
  \BibitemOpen
  \bibfield  {author} {\bibinfo {author} {\bibfnamefont {J.~J.~L.}\
  \bibnamefont {Morton}}, \bibinfo {author} {\bibfnamefont {D.~R.}\
  \bibnamefont {McCamey}}, \bibinfo {author} {\bibfnamefont {M.~A.}\
  \bibnamefont {Eriksson}},\ and\ \bibinfo {author} {\bibfnamefont {S.~A.}\
  \bibnamefont {Lyon}},\ }\href {https://doi.org/10.1038/nature10681}
  {\bibfield  {journal} {\bibinfo  {journal} {Nature}\ }\textbf {\bibinfo
  {volume} {479}},\ \bibinfo {pages} {345} (\bibinfo {year}
  {2011})}\BibitemShut {NoStop}%
\bibitem [{\citenamefont {Kadykov}\ \emph
  {et~al.}(2016{\natexlab{b}})\citenamefont {Kadykov}, \citenamefont {Torres},
  \citenamefont {Krishtopenko}, \citenamefont {Consejo}, \citenamefont
  {Ruffenach}, \citenamefont {Marcinkiewicz}, \citenamefont {But},
  \citenamefont {Knap}, \citenamefont {Morozov}, \citenamefont {Gavrilenko},
  \citenamefont {Mikhailov}, \citenamefont {Dvoretsky},\ and\ \citenamefont
  {Teppe}}]{kadykovTerahertzImagingLandau2016b}%
  \BibitemOpen
  \bibfield  {author} {\bibinfo {author} {\bibfnamefont {A.~M.}\ \bibnamefont
  {Kadykov}}, \bibinfo {author} {\bibfnamefont {J.}~\bibnamefont {Torres}},
  \bibinfo {author} {\bibfnamefont {S.~S.}\ \bibnamefont {Krishtopenko}},
  \bibinfo {author} {\bibfnamefont {C.}~\bibnamefont {Consejo}}, \bibinfo
  {author} {\bibfnamefont {S.}~\bibnamefont {Ruffenach}}, \bibinfo {author}
  {\bibfnamefont {M.}~\bibnamefont {Marcinkiewicz}}, \bibinfo {author}
  {\bibfnamefont {D.}~\bibnamefont {But}}, \bibinfo {author} {\bibfnamefont
  {W.}~\bibnamefont {Knap}}, \bibinfo {author} {\bibfnamefont {S.~V.}\
  \bibnamefont {Morozov}}, \bibinfo {author} {\bibfnamefont {V.~I.}\
  \bibnamefont {Gavrilenko}}, \bibinfo {author} {\bibfnamefont {N.~N.}\
  \bibnamefont {Mikhailov}}, \bibinfo {author} {\bibfnamefont {S.~A.}\
  \bibnamefont {Dvoretsky}},\ and\ \bibinfo {author} {\bibfnamefont
  {F.}~\bibnamefont {Teppe}},\ }\href {https://doi.org/10.1063/1.4955018}
  {\bibfield  {journal} {\bibinfo  {journal} {Applied Physics Letters}\
  }\textbf {\bibinfo {volume} {108}},\ \bibinfo {pages} {262102} (\bibinfo
  {year} {2016}{\natexlab{b}})}\BibitemShut {NoStop}%
\bibitem [{\citenamefont {Videlier}\ \emph {et~al.}(2011)\citenamefont
  {Videlier}, \citenamefont {Dyakonova}, \citenamefont {Teppe}, \citenamefont
  {Consejo}, \citenamefont {Chenaud}, \citenamefont {Knap}, \citenamefont
  {Lusakowski}, \citenamefont {Tomaszewski}, \citenamefont {Marczewski},\ and\
  \citenamefont {Grabiec}}]{videlierTerahertzPhotovoltaicResponse2011a}%
  \BibitemOpen
  \bibfield  {author} {\bibinfo {author} {\bibfnamefont {H.}~\bibnamefont
  {Videlier}}, \bibinfo {author} {\bibfnamefont {N.}~\bibnamefont {Dyakonova}},
  \bibinfo {author} {\bibfnamefont {F.}~\bibnamefont {Teppe}}, \bibinfo
  {author} {\bibfnamefont {C.}~\bibnamefont {Consejo}}, \bibinfo {author}
  {\bibfnamefont {B.}~\bibnamefont {Chenaud}}, \bibinfo {author} {\bibfnamefont
  {W.}~\bibnamefont {Knap}}, \bibinfo {author} {\bibfnamefont {J.}~\bibnamefont
  {Lusakowski}}, \bibinfo {author} {\bibfnamefont {D.}~\bibnamefont
  {Tomaszewski}}, \bibinfo {author} {\bibfnamefont {J.}~\bibnamefont
  {Marczewski}},\ and\ \bibinfo {author} {\bibfnamefont {P.}~\bibnamefont
  {Grabiec}},\ }\href {https://doi.org/10.12693/APhysPolA.120.927} {\bibfield
  {journal} {\bibinfo  {journal} {Acta Physica Polonica A}\ }\textbf {\bibinfo
  {volume} {120}},\ \bibinfo {pages} {927} (\bibinfo {year}
  {2011})}\BibitemShut {NoStop}%
\bibitem [{\citenamefont
  {Lepine}(1972)}]{lepineSpinDependentRecombinationSilicon1972a}%
  \BibitemOpen
  \bibfield  {author} {\bibinfo {author} {\bibfnamefont {D.~J.}\ \bibnamefont
  {Lepine}},\ }\href {https://doi.org/10.1103/PhysRevB.6.436} {\bibfield
  {journal} {\bibinfo  {journal} {Physical Review B}\ }\textbf {\bibinfo
  {volume} {6}},\ \bibinfo {pages} {436} (\bibinfo {year} {1972})}\BibitemShut
  {NoStop}%
\bibitem [{\citenamefont {Solomon}\ \emph {et~al.}(1977)\citenamefont
  {Solomon}, \citenamefont {Biegelsen},\ and\ \citenamefont
  {Knights}}]{solomonSpindependentPhotoconductivityNtype1977a}%
  \BibitemOpen
  \bibfield  {author} {\bibinfo {author} {\bibfnamefont {I.}~\bibnamefont
  {Solomon}}, \bibinfo {author} {\bibfnamefont {D.}~\bibnamefont {Biegelsen}},\
  and\ \bibinfo {author} {\bibfnamefont {J.~C.}\ \bibnamefont {Knights}},\
  }\href {https://doi.org/10.1016/0038-1098(77)91402-8} {\bibfield  {journal}
  {\bibinfo  {journal} {Solid State Communications}\ }\textbf {\bibinfo
  {volume} {22}},\ \bibinfo {pages} {505} (\bibinfo {year} {1977})}\BibitemShut
  {NoStop}%
\bibitem [{\citenamefont {Dersch}\ \emph {et~al.}(1983)\citenamefont {Dersch},
  \citenamefont {Schweitzer},\ and\ \citenamefont
  {Stuke}}]{derschRecombinationProcesses$a$Si1983}%
  \BibitemOpen
  \bibfield  {author} {\bibinfo {author} {\bibfnamefont {H.}~\bibnamefont
  {Dersch}}, \bibinfo {author} {\bibfnamefont {L.}~\bibnamefont {Schweitzer}},\
  and\ \bibinfo {author} {\bibfnamefont {J.}~\bibnamefont {Stuke}},\ }\href
  {https://doi.org/10.1103/PhysRevB.28.4678} {\bibfield  {journal} {\bibinfo
  {journal} {Physical Review B}\ }\textbf {\bibinfo {volume} {28}},\ \bibinfo
  {pages} {4678} (\bibinfo {year} {1983})}\BibitemShut {NoStop}%
\bibitem [{\citenamefont
  {Street}(1982)}]{streetSpindependentPhotoconductivityUndoped1982}%
  \BibitemOpen
  \bibfield  {author} {\bibinfo {author} {\bibfnamefont {R.~A.}\ \bibnamefont
  {Street}},\ }\href {https://doi.org/10.1080/13642818208246439} {\bibfield
  {journal} {\bibinfo  {journal} {Philosophical Magazine B}\ }\textbf {\bibinfo
  {volume} {46}},\ \bibinfo {pages} {273} (\bibinfo {year} {1982})}\BibitemShut
  {NoStop}%
\bibitem [{\citenamefont {Schiff}(1981)}]{schiffSpinPolarizationEffects1981}%
  \BibitemOpen
  \bibfield  {author} {\bibinfo {author} {\bibfnamefont {E.~A.}\ \bibnamefont
  {Schiff}},\ }\href {https://doi.org/10.1063/1.33036} {\bibfield  {journal}
  {\bibinfo  {journal} {AIP Conference Proceedings}\ }\textbf {\bibinfo
  {volume} {73}},\ \bibinfo {pages} {233} (\bibinfo {year} {1981})}\BibitemShut
  {NoStop}%
\bibitem [{\citenamefont {Kaplan}\ \emph {et~al.}(1978)\citenamefont {Kaplan},
  \citenamefont {Solomon},\ and\ \citenamefont
  {Mott}}]{kaplanExplanationLargeSpindependent1978a}%
  \BibitemOpen
  \bibfield  {author} {\bibinfo {author} {\bibfnamefont {D.}~\bibnamefont
  {Kaplan}}, \bibinfo {author} {\bibfnamefont {I.}~\bibnamefont {Solomon}},\
  and\ \bibinfo {author} {\bibfnamefont {N.~F.}\ \bibnamefont {Mott}},\ }\href
  {https://doi.org/10.1051/jphyslet:0197800390405100} {\bibfield  {journal}
  {\bibinfo  {journal} {Journal de Physique Lettres}\ }\textbf {\bibinfo
  {volume} {39}},\ \bibinfo {pages} {51} (\bibinfo {year} {1978})}\BibitemShut
  {NoStop}%
\bibitem [{\citenamefont {Bhatt}\ \emph {et~al.}(2022)\citenamefont {Bhatt},
  \citenamefont {Kim},\ and\ \citenamefont
  {Kim}}]{bhattVariousDefectsGraphene2022}%
  \BibitemOpen
  \bibfield  {author} {\bibinfo {author} {\bibfnamefont {M.~D.}\ \bibnamefont
  {Bhatt}}, \bibinfo {author} {\bibfnamefont {H.}~\bibnamefont {Kim}},\ and\
  \bibinfo {author} {\bibfnamefont {G.}~\bibnamefont {Kim}},\ }\href
  {https://doi.org/10.1039/D2RA01436J} {\bibfield  {journal} {\bibinfo
  {journal} {RSC Advances}\ }\textbf {\bibinfo {volume} {12}},\ \bibinfo
  {pages} {21520} (\bibinfo {year} {2022})}\BibitemShut {NoStop}%
\bibitem [{\citenamefont {Abid}\ \emph {et~al.}(2018)\citenamefont {Abid},
  \citenamefont {Sehrawat}, \citenamefont {Islam}, \citenamefont {Mishra},\
  and\ \citenamefont {Ahmad}}]{abidReducedGrapheneOxide2018}%
  \BibitemOpen
  \bibfield  {author} {\bibinfo {author} {\bibnamefont {Abid}}, \bibinfo
  {author} {\bibfnamefont {P.}~\bibnamefont {Sehrawat}}, \bibinfo {author}
  {\bibfnamefont {S.~S.}\ \bibnamefont {Islam}}, \bibinfo {author}
  {\bibfnamefont {P.}~\bibnamefont {Mishra}},\ and\ \bibinfo {author}
  {\bibfnamefont {S.}~\bibnamefont {Ahmad}},\ }\href
  {https://doi.org/10.1038/s41598-018-21686-2} {\bibfield  {journal} {\bibinfo
  {journal} {Scientific Reports}\ }\textbf {\bibinfo {volume} {8}},\ \bibinfo
  {pages} {3537} (\bibinfo {year} {2018})}\BibitemShut {NoStop}%
\bibitem [{\citenamefont {Dutta}\ and\ \citenamefont
  {Wakabayashi}(2015)}]{duttaMagnetizationDueLocalized2015}%
  \BibitemOpen
  \bibfield  {author} {\bibinfo {author} {\bibfnamefont {S.}~\bibnamefont
  {Dutta}}\ and\ \bibinfo {author} {\bibfnamefont {K.}~\bibnamefont
  {Wakabayashi}},\ }\href {https://doi.org/10.1038/srep11744} {\bibfield
  {journal} {\bibinfo  {journal} {Scientific Reports}\ }\textbf {\bibinfo
  {volume} {5}},\ \bibinfo {pages} {11744} (\bibinfo {year}
  {2015})}\BibitemShut {NoStop}%
\bibitem [{\citenamefont {Ghising}\ \emph {et~al.}(2023)\citenamefont
  {Ghising}, \citenamefont {Biswas},\ and\ \citenamefont
  {Lee}}]{ghisingGrapheneSpinValves2023}%
  \BibitemOpen
  \bibfield  {author} {\bibinfo {author} {\bibfnamefont {P.}~\bibnamefont
  {Ghising}}, \bibinfo {author} {\bibfnamefont {C.}~\bibnamefont {Biswas}},\
  and\ \bibinfo {author} {\bibfnamefont {Y.~H.}\ \bibnamefont {Lee}},\ }\href
  {https://doi.org/10.1002/adma.202209137} {\bibfield  {journal} {\bibinfo
  {journal} {Advanced Materials}\ }\textbf {\bibinfo {volume} {35}},\ \bibinfo
  {pages} {2209137} (\bibinfo {year} {2023})}\BibitemShut {NoStop}%
\bibitem [{\citenamefont {Cavenett}\ and\ \citenamefont
  {Brunwin}(1979)}]{cavenettSpinDependentConductivity1979}%
  \BibitemOpen
  \bibfield  {author} {\bibinfo {author} {\bibfnamefont {B.~C.}\ \bibnamefont
  {Cavenett}}\ and\ \bibinfo {author} {\bibfnamefont {R.~F.}\ \bibnamefont
  {Brunwin}},\ }\href {https://doi.org/10.1016/0038-1098(79)90318-1} {\bibfield
   {journal} {\bibinfo  {journal} {Solid State Communications}\ }\textbf
  {\bibinfo {volume} {31}},\ \bibinfo {pages} {659} (\bibinfo {year}
  {1979})}\BibitemShut {NoStop}%
\bibitem [{\citenamefont {{de Sousa}}\ \emph {et~al.}(2009)\citenamefont {{de
  Sousa}}, \citenamefont {Lo},\ and\ \citenamefont
  {Bokor}}]{desousaSpindependentScatteringSilicon2009}%
  \BibitemOpen
  \bibfield  {author} {\bibinfo {author} {\bibfnamefont {R.}~\bibnamefont {{de
  Sousa}}}, \bibinfo {author} {\bibfnamefont {C.~C.}\ \bibnamefont {Lo}},\ and\
  \bibinfo {author} {\bibfnamefont {J.}~\bibnamefont {Bokor}},\ }\href
  {https://doi.org/10.1103/PhysRevB.80.045320} {\bibfield  {journal} {\bibinfo
  {journal} {Physical Review B}\ }\textbf {\bibinfo {volume} {80}},\ \bibinfo
  {pages} {045320} (\bibinfo {year} {2009})}\BibitemShut {NoStop}%
\bibitem [{\citenamefont {{L. R.}}(1961)}]{l.r.BookReviewPrinciples1961}%
  \BibitemOpen
  \bibfield  {author} {\bibinfo {author} {\bibnamefont {{L. R.}}},\ }\href
  {https://doi.org/10.1016/0029-5582(61)90091-8} {\bibfield  {journal}
  {\bibinfo  {journal} {Nuclear Physics}\ }\textbf {\bibinfo {volume} {28}},\
  \bibinfo {pages} {692} (\bibinfo {year} {1961})}\BibitemShut {NoStop}%
\bibitem [{\citenamefont {Weil}\ and\ \citenamefont
  {Bolton}(2007)}]{weilElectronParamagneticResonance2007}%
  \BibitemOpen
  \bibfield  {author} {\bibinfo {author} {\bibfnamefont {J.~A.}\ \bibnamefont
  {Weil}}\ and\ \bibinfo {author} {\bibfnamefont {J.~R.}\ \bibnamefont
  {Bolton}},\ }\href@noop {} {\emph {\bibinfo {title} {Electron {{Paramagnetic
  Resonance}}: {{Elementary Theory}} and {{Practical Applications}}}}}\
  (\bibinfo  {publisher} {John Wiley \& Sons},\ \bibinfo {year}
  {2007})\BibitemShut {NoStop}%
\bibitem [{dat(2025)}]{dataset}%
  \BibitemOpen
  \href@noop {} {\bibinfo {title} {Data repository}},\ \bibinfo {howpublished}
  {\url{https://doi.org/10.5281/zenodo.15098846}} (\bibinfo {year}
  {2025})\BibitemShut {NoStop}%
\bibitem [{\citenamefont {Knap}\ \emph {et~al.}(2002)\citenamefont {Knap},
  \citenamefont {Kachorovskii}, \citenamefont {Deng}, \citenamefont
  {Rumyantsev}, \citenamefont {L{\"u}}, \citenamefont {Gaska}, \citenamefont
  {Shur}, \citenamefont {Simin}, \citenamefont {Hu}, \citenamefont {Khan},
  \citenamefont {Saylor},\ and\ \citenamefont
  {Brunel}}]{knapNonresonantDetectionTerahertz2002}%
  \BibitemOpen
  \bibfield  {author} {\bibinfo {author} {\bibfnamefont {W.}~\bibnamefont
  {Knap}}, \bibinfo {author} {\bibfnamefont {V.}~\bibnamefont {Kachorovskii}},
  \bibinfo {author} {\bibfnamefont {Y.}~\bibnamefont {Deng}}, \bibinfo {author}
  {\bibfnamefont {S.}~\bibnamefont {Rumyantsev}}, \bibinfo {author}
  {\bibfnamefont {J.-Q.}\ \bibnamefont {L{\"u}}}, \bibinfo {author}
  {\bibfnamefont {R.}~\bibnamefont {Gaska}}, \bibinfo {author} {\bibfnamefont
  {M.~S.}\ \bibnamefont {Shur}}, \bibinfo {author} {\bibfnamefont
  {G.}~\bibnamefont {Simin}}, \bibinfo {author} {\bibfnamefont
  {X.}~\bibnamefont {Hu}}, \bibinfo {author} {\bibfnamefont {M.~A.}\
  \bibnamefont {Khan}}, \bibinfo {author} {\bibfnamefont {C.~A.}\ \bibnamefont
  {Saylor}},\ and\ \bibinfo {author} {\bibfnamefont {L.~C.}\ \bibnamefont
  {Brunel}},\ }\href {https://doi.org/10.1063/1.1468257} {\bibfield  {journal}
  {\bibinfo  {journal} {Journal of Applied Physics}\ }\textbf {\bibinfo
  {volume} {91}},\ \bibinfo {pages} {9346} (\bibinfo {year}
  {2002})}\BibitemShut {NoStop}%
\bibitem [{\citenamefont {Wang}\ \emph {et~al.}(2013)\citenamefont {Wang},
  \citenamefont {Meric}, \citenamefont {Huang}, \citenamefont {Gao},
  \citenamefont {Gao}, \citenamefont {Tran}, \citenamefont {Taniguchi},
  \citenamefont {Watanabe}, \citenamefont {Campos}, \citenamefont {Muller},
  \citenamefont {Guo}, \citenamefont {Kim}, \citenamefont {Hone}, \citenamefont
  {Shepard},\ and\ \citenamefont
  {Dean}}]{wangOneDimensionalElectricalContact2013a}%
  \BibitemOpen
  \bibfield  {author} {\bibinfo {author} {\bibfnamefont {L.}~\bibnamefont
  {Wang}}, \bibinfo {author} {\bibfnamefont {I.}~\bibnamefont {Meric}},
  \bibinfo {author} {\bibfnamefont {P.~Y.}\ \bibnamefont {Huang}}, \bibinfo
  {author} {\bibfnamefont {Q.}~\bibnamefont {Gao}}, \bibinfo {author}
  {\bibfnamefont {Y.}~\bibnamefont {Gao}}, \bibinfo {author} {\bibfnamefont
  {H.}~\bibnamefont {Tran}}, \bibinfo {author} {\bibfnamefont {T.}~\bibnamefont
  {Taniguchi}}, \bibinfo {author} {\bibfnamefont {K.}~\bibnamefont {Watanabe}},
  \bibinfo {author} {\bibfnamefont {L.~M.}\ \bibnamefont {Campos}}, \bibinfo
  {author} {\bibfnamefont {D.~A.}\ \bibnamefont {Muller}}, \bibinfo {author}
  {\bibfnamefont {J.}~\bibnamefont {Guo}}, \bibinfo {author} {\bibfnamefont
  {P.}~\bibnamefont {Kim}}, \bibinfo {author} {\bibfnamefont {J.}~\bibnamefont
  {Hone}}, \bibinfo {author} {\bibfnamefont {K.~L.}\ \bibnamefont {Shepard}},\
  and\ \bibinfo {author} {\bibfnamefont {C.~R.}\ \bibnamefont {Dean}},\ }\href
  {https://doi.org/10.1126/science.1244358} {\bibfield  {journal} {\bibinfo
  {journal} {Science}\ }\textbf {\bibinfo {volume} {342}},\ \bibinfo {pages}
  {614} (\bibinfo {year} {2013})}\BibitemShut {NoStop}%
\bibitem [{\citenamefont {Moczko}\ \emph {et~al.}(2023)\citenamefont {Moczko},
  \citenamefont {Reichardt}, \citenamefont {Singh}, \citenamefont {Zhang},
  \citenamefont {L{\'o}pez}, \citenamefont {Wolff}, \citenamefont {Moghe},
  \citenamefont {Lorchat}, \citenamefont {Singh}, \citenamefont {Watanabe},
  \citenamefont {Taniguchi}, \citenamefont {Majjad}, \citenamefont {Romeo},
  \citenamefont {Gloppe}, \citenamefont {Wirtz},\ and\ \citenamefont
  {Berciaud}}]{moczkoSymmetrydependentDielectricScreening2023a}%
  \BibitemOpen
  \bibfield  {author} {\bibinfo {author} {\bibfnamefont {L.}~\bibnamefont
  {Moczko}}, \bibinfo {author} {\bibfnamefont {S.}~\bibnamefont {Reichardt}},
  \bibinfo {author} {\bibfnamefont {A.}~\bibnamefont {Singh}}, \bibinfo
  {author} {\bibfnamefont {X.}~\bibnamefont {Zhang}}, \bibinfo {author}
  {\bibfnamefont {L.~E.~P.}\ \bibnamefont {L{\'o}pez}}, \bibinfo {author}
  {\bibfnamefont {J.~L.~P.}\ \bibnamefont {Wolff}}, \bibinfo {author}
  {\bibfnamefont {A.~R.}\ \bibnamefont {Moghe}}, \bibinfo {author}
  {\bibfnamefont {E.}~\bibnamefont {Lorchat}}, \bibinfo {author} {\bibfnamefont
  {R.}~\bibnamefont {Singh}}, \bibinfo {author} {\bibfnamefont
  {K.}~\bibnamefont {Watanabe}}, \bibinfo {author} {\bibfnamefont
  {T.}~\bibnamefont {Taniguchi}}, \bibinfo {author} {\bibfnamefont
  {H.}~\bibnamefont {Majjad}}, \bibinfo {author} {\bibfnamefont
  {M.}~\bibnamefont {Romeo}}, \bibinfo {author} {\bibfnamefont
  {A.}~\bibnamefont {Gloppe}}, \bibinfo {author} {\bibfnamefont
  {L.}~\bibnamefont {Wirtz}},\ and\ \bibinfo {author} {\bibfnamefont
  {S.}~\bibnamefont {Berciaud}},\ }\href
  {https://doi.org/10.48550/arXiv.2310.13868} {\bibinfo {title}
  {Symmetry-dependent dielectric screening of optical phonons in monolayer
  graphene}} (\bibinfo {year} {2023}),\ \Eprint
  {https://arxiv.org/abs/2310.13868} {arXiv:2310.13868 [cond-mat]} \BibitemShut
  {NoStop}%
\bibitem [{\citenamefont {Viti}\ \emph {et~al.}(2016)\citenamefont {Viti},
  \citenamefont {Hu}, \citenamefont {Coquillat}, \citenamefont {Politano},
  \citenamefont {Knap},\ and\ \citenamefont
  {Vitiello}}]{vitiEfficientTerahertzDetection2016}%
  \BibitemOpen
  \bibfield  {author} {\bibinfo {author} {\bibfnamefont {L.}~\bibnamefont
  {Viti}}, \bibinfo {author} {\bibfnamefont {J.}~\bibnamefont {Hu}}, \bibinfo
  {author} {\bibfnamefont {D.}~\bibnamefont {Coquillat}}, \bibinfo {author}
  {\bibfnamefont {A.}~\bibnamefont {Politano}}, \bibinfo {author}
  {\bibfnamefont {W.}~\bibnamefont {Knap}},\ and\ \bibinfo {author}
  {\bibfnamefont {M.~S.}\ \bibnamefont {Vitiello}},\ }\href
  {https://doi.org/10.1038/srep20474} {\bibfield  {journal} {\bibinfo
  {journal} {Scientific Reports}\ }\textbf {\bibinfo {volume} {6}},\ \bibinfo
  {pages} {20474} (\bibinfo {year} {2016})}\BibitemShut {NoStop}%
\bibitem [{\citenamefont {Shabanov}\ \emph {et~al.}(2021)\citenamefont
  {Shabanov}, \citenamefont {Moskotin}, \citenamefont {Belosevich},
  \citenamefont {Matyushkin}, \citenamefont {Rybin}, \citenamefont {Fedorov},\
  and\ \citenamefont {Svintsov}}]{shabanovOptimalAsymmetryTransistorbased2021}%
  \BibitemOpen
  \bibfield  {author} {\bibinfo {author} {\bibfnamefont {A.}~\bibnamefont
  {Shabanov}}, \bibinfo {author} {\bibfnamefont {M.}~\bibnamefont {Moskotin}},
  \bibinfo {author} {\bibfnamefont {V.}~\bibnamefont {Belosevich}}, \bibinfo
  {author} {\bibfnamefont {Y.}~\bibnamefont {Matyushkin}}, \bibinfo {author}
  {\bibfnamefont {M.}~\bibnamefont {Rybin}}, \bibinfo {author} {\bibfnamefont
  {G.}~\bibnamefont {Fedorov}},\ and\ \bibinfo {author} {\bibfnamefont
  {D.}~\bibnamefont {Svintsov}},\ }\href {https://doi.org/10.1063/5.0063870}
  {\bibfield  {journal} {\bibinfo  {journal} {Applied Physics Letters}\
  }\textbf {\bibinfo {volume} {119}},\ \bibinfo {pages} {163505} (\bibinfo
  {year} {2021})}\BibitemShut {NoStop}%
\bibitem [{\citenamefont {Cai}\ \emph {et~al.}(2014)\citenamefont {Cai},
  \citenamefont {Sushkov}, \citenamefont {Suess}, \citenamefont {Jadidi},
  \citenamefont {Jenkins}, \citenamefont {Nyakiti}, \citenamefont
  {{Myers-Ward}}, \citenamefont {Li}, \citenamefont {Yan}, \citenamefont
  {Gaskill}, \citenamefont {Murphy}, \citenamefont {Drew},\ and\ \citenamefont
  {Fuhrer}}]{caiSensitiveRoomtemperatureTerahertz2014}%
  \BibitemOpen
  \bibfield  {author} {\bibinfo {author} {\bibfnamefont {X.}~\bibnamefont
  {Cai}}, \bibinfo {author} {\bibfnamefont {A.~B.}\ \bibnamefont {Sushkov}},
  \bibinfo {author} {\bibfnamefont {R.~J.}\ \bibnamefont {Suess}}, \bibinfo
  {author} {\bibfnamefont {M.~M.}\ \bibnamefont {Jadidi}}, \bibinfo {author}
  {\bibfnamefont {G.~S.}\ \bibnamefont {Jenkins}}, \bibinfo {author}
  {\bibfnamefont {L.~O.}\ \bibnamefont {Nyakiti}}, \bibinfo {author}
  {\bibfnamefont {R.~L.}\ \bibnamefont {{Myers-Ward}}}, \bibinfo {author}
  {\bibfnamefont {S.}~\bibnamefont {Li}}, \bibinfo {author} {\bibfnamefont
  {J.}~\bibnamefont {Yan}}, \bibinfo {author} {\bibfnamefont {D.~K.}\
  \bibnamefont {Gaskill}}, \bibinfo {author} {\bibfnamefont {T.~E.}\
  \bibnamefont {Murphy}}, \bibinfo {author} {\bibfnamefont {H.~D.}\
  \bibnamefont {Drew}},\ and\ \bibinfo {author} {\bibfnamefont {M.~S.}\
  \bibnamefont {Fuhrer}},\ }\href {https://doi.org/10.1038/nnano.2014.182}
  {\bibfield  {journal} {\bibinfo  {journal} {Nature Nanotechnology}\ }\textbf
  {\bibinfo {volume} {9}},\ \bibinfo {pages} {814} (\bibinfo {year}
  {2014})}\BibitemShut {NoStop}%
\bibitem [{\citenamefont {Ludwig}\ \emph {et~al.}(2024)\citenamefont {Ludwig},
  \citenamefont {Generalov}, \citenamefont {Holstein}, \citenamefont {Murros},
  \citenamefont {Viisanen}, \citenamefont {Prunnila},\ and\ \citenamefont
  {Roskos}}]{ludwigTerahertzDetectionGraphene2024}%
  \BibitemOpen
  \bibfield  {author} {\bibinfo {author} {\bibfnamefont {F.}~\bibnamefont
  {Ludwig}}, \bibinfo {author} {\bibfnamefont {A.}~\bibnamefont {Generalov}},
  \bibinfo {author} {\bibfnamefont {J.}~\bibnamefont {Holstein}}, \bibinfo
  {author} {\bibfnamefont {A.}~\bibnamefont {Murros}}, \bibinfo {author}
  {\bibfnamefont {K.}~\bibnamefont {Viisanen}}, \bibinfo {author}
  {\bibfnamefont {M.}~\bibnamefont {Prunnila}},\ and\ \bibinfo {author}
  {\bibfnamefont {H.~G.}\ \bibnamefont {Roskos}},\ }\href
  {https://doi.org/10.1021/acsaelm.3c01511} {\bibfield  {journal} {\bibinfo
  {journal} {ACS Applied Electronic Materials}\ }\textbf {\bibinfo {volume}
  {6}},\ \bibinfo {pages} {2197} (\bibinfo {year} {2024})}\BibitemShut
  {NoStop}%
\bibitem [{\citenamefont {Low}\ \emph {et~al.}(2014)\citenamefont {Low},
  \citenamefont {Engel}, \citenamefont {Steiner},\ and\ \citenamefont
  {Avouris}}]{lowOriginPhotoresponseBlack2014}%
  \BibitemOpen
  \bibfield  {author} {\bibinfo {author} {\bibfnamefont {T.}~\bibnamefont
  {Low}}, \bibinfo {author} {\bibfnamefont {M.}~\bibnamefont {Engel}}, \bibinfo
  {author} {\bibfnamefont {M.}~\bibnamefont {Steiner}},\ and\ \bibinfo {author}
  {\bibfnamefont {P.}~\bibnamefont {Avouris}},\ }\href
  {https://doi.org/10.1103/PhysRevB.90.081408} {\bibfield  {journal} {\bibinfo
  {journal} {Physical Review B}\ }\textbf {\bibinfo {volume} {90}},\ \bibinfo
  {pages} {081408} (\bibinfo {year} {2014})}\BibitemShut {NoStop}%
\bibitem [{\citenamefont {Skoblin}\ \emph {et~al.}(2017)\citenamefont
  {Skoblin}, \citenamefont {Sun},\ and\ \citenamefont
  {Yurgens}}]{skoblinThermoelectricEffectsGraphene2017}%
  \BibitemOpen
  \bibfield  {author} {\bibinfo {author} {\bibfnamefont {G.}~\bibnamefont
  {Skoblin}}, \bibinfo {author} {\bibfnamefont {J.}~\bibnamefont {Sun}},\ and\
  \bibinfo {author} {\bibfnamefont {A.}~\bibnamefont {Yurgens}},\ }\href
  {https://doi.org/10.1038/s41598-017-15857-w} {\bibfield  {journal} {\bibinfo
  {journal} {Scientific Reports}\ }\textbf {\bibinfo {volume} {7}},\ \bibinfo
  {pages} {15542} (\bibinfo {year} {2017})}\BibitemShut {NoStop}%
\bibitem [{\citenamefont {Harzheim}\ \emph {et~al.}(2018)\citenamefont
  {Harzheim}, \citenamefont {Spiece}, \citenamefont {Evangeli}, \citenamefont
  {McCann}, \citenamefont {Falko}, \citenamefont {Sheng}, \citenamefont
  {Warner}, \citenamefont {Briggs}, \citenamefont {Mol}, \citenamefont
  {Gehring},\ and\ \citenamefont
  {Kolosov}}]{harzheimGeometricallyEnhancedThermoelectric2018}%
  \BibitemOpen
  \bibfield  {author} {\bibinfo {author} {\bibfnamefont {A.}~\bibnamefont
  {Harzheim}}, \bibinfo {author} {\bibfnamefont {J.}~\bibnamefont {Spiece}},
  \bibinfo {author} {\bibfnamefont {C.}~\bibnamefont {Evangeli}}, \bibinfo
  {author} {\bibfnamefont {E.}~\bibnamefont {McCann}}, \bibinfo {author}
  {\bibfnamefont {V.}~\bibnamefont {Falko}}, \bibinfo {author} {\bibfnamefont
  {Y.}~\bibnamefont {Sheng}}, \bibinfo {author} {\bibfnamefont {J.~H.}\
  \bibnamefont {Warner}}, \bibinfo {author} {\bibfnamefont {G.~A.~D.}\
  \bibnamefont {Briggs}}, \bibinfo {author} {\bibfnamefont {J.~A.}\
  \bibnamefont {Mol}}, \bibinfo {author} {\bibfnamefont {P.}~\bibnamefont
  {Gehring}},\ and\ \bibinfo {author} {\bibfnamefont {O.~V.}\ \bibnamefont
  {Kolosov}},\ }\href {https://doi.org/10.1021/acs.nanolett.8b03406} {\bibfield
   {journal} {\bibinfo  {journal} {Nano Letters}\ }\textbf {\bibinfo {volume}
  {18}},\ \bibinfo {pages} {7719} (\bibinfo {year} {2018})}\BibitemShut
  {NoStop}%
\bibitem [{\citenamefont {Anno}\ \emph {et~al.}(2017)\citenamefont {Anno},
  \citenamefont {Imakita}, \citenamefont {Takei}, \citenamefont {Akita},\ and\
  \citenamefont {Arie}}]{annoEnhancementGrapheneThermoelectric2017}%
  \BibitemOpen
  \bibfield  {author} {\bibinfo {author} {\bibfnamefont {Y.}~\bibnamefont
  {Anno}}, \bibinfo {author} {\bibfnamefont {Y.}~\bibnamefont {Imakita}},
  \bibinfo {author} {\bibfnamefont {K.}~\bibnamefont {Takei}}, \bibinfo
  {author} {\bibfnamefont {S.}~\bibnamefont {Akita}},\ and\ \bibinfo {author}
  {\bibfnamefont {T.}~\bibnamefont {Arie}},\ }\href
  {https://doi.org/10.1088/2053-1583/aa57fc} {\bibfield  {journal} {\bibinfo
  {journal} {2D Materials}\ }\textbf {\bibinfo {volume} {4}},\ \bibinfo {pages}
  {025019} (\bibinfo {year} {2017})}\BibitemShut {NoStop}%
\bibitem [{\citenamefont {Dollfus}\ \emph {et~al.}(2015)\citenamefont
  {Dollfus}, \citenamefont {Hung~Nguyen},\ and\ \citenamefont
  {{Saint-Martin}}}]{dollfusThermoelectricEffectsGraphene2015}%
  \BibitemOpen
  \bibfield  {author} {\bibinfo {author} {\bibfnamefont {P.}~\bibnamefont
  {Dollfus}}, \bibinfo {author} {\bibfnamefont {V.}~\bibnamefont
  {Hung~Nguyen}},\ and\ \bibinfo {author} {\bibfnamefont {J.}~\bibnamefont
  {{Saint-Martin}}},\ }\href {https://doi.org/10.1088/0953-8984/27/13/133204}
  {\bibfield  {journal} {\bibinfo  {journal} {Journal of Physics: Condensed
  Matter}\ }\textbf {\bibinfo {volume} {27}},\ \bibinfo {pages} {133204}
  (\bibinfo {year} {2015})}\BibitemShut {NoStop}%
\end{thebibliography}%

\end{document}